\documentclass[proof]{pasj00}
\usepackage{array}
\usepackage{url}
\SetRunningHead{K. Sakai et al.}{Structual Study of Galactic Hot Gas toward Mkn 421}

\begin{document}

\title{Structual Study of Galactic Hot Gas toward Markarian 421 from X-Ray Absorption and Emission Lines}

\author{Kazuhiro \textsc{Sakai},\altaffilmark{1} Yangsen
\textsc{Yao},\altaffilmark{2} Kazuhisa
\textsc{Mitsuda},\altaffilmark{1} Noriko \textsc{Yamasaki},\altaffilmark{1} Q.
Daniel \textsc{Wang},\altaffilmark{3} Yoh \textsc{Takei},\altaffilmark{1} \\
and Dan \textsc{McCammon}\altaffilmark{4}}

\altaffiltext{1}{Institute of Space and Astronautical Science, Japan
  Aerospace Exploration Agency, 3-1-1 Yoshinodai, Chuo, Sagamihara,
  Kanagawa 252-5210, Japan}
\altaffiltext{2}{Eureka Scientific, Inc., 2452 Delmer Street Suite 100,
  Oakland, CA 94602, USA} 
\altaffiltext{3}{Department of Astronomy,
  University of Massachusetts, Amherst, MA 01003, USA}
\altaffiltext{4}{Department of Physics, University of Wisconsin,
  Madison, 1150 University Avenue, Madison, WI 53706, USA}

\email{sakai@astro.isas.jaxa.jp}

\maketitle

\begin{abstract}
  We present a structural study of the hot ISM in the Galactic halo
  along the sight line toward the bright active galactic nucleus Mkn
  421. The O\emissiontype{VII} and O\emissiontype{VIII} absorption
  lines were measured with the Low Energy Transmission Grating
  Spectrograph aboard \emph{Chandra} toward Mkn 421, and the
  O\emissiontype{VII} and O\emissiontype{VIII} emission lines were
  observed in the adjacent fields of the sight line with the X-ray
  Imaging Spectrometer aboard \emph{Suzaku}. We jointly analyzed the
  absorption and the emission spectra assuming exponential
  distributions of the gas temperature and density from the Galactic
  plane, and constrained the temperature and density at the plane to
  be $(3.2^{+0.6}_{-0.7})\times 10^6\,\mathrm{K}$ and
  $(1.2^{+0.5}_{-0.4})\times 10^{-3}\,\mathrm{cm^{-3}}$, with the
  scale heights of $1.6^{+1.7}_{-0.7}\,\mathrm{kpc}$ and
  $>2.8\,\mathrm{kpc}$ respectively. The results are consistent with
  those obtained in the LMC X--3 direction and the PKS 2155--304
  direction, describing a thick disk-like hot gas with its height of a
  few kpc from the Galactic plane.
\end{abstract}

\section{Introduction}

The observation of the whole sky in the soft X-ray band, especially
between 0.5 and 1 keV, has clearly showed not only enhanced emission
from regions such as the Galactic Center and the North Polar Spur, but
also a dimmer yet gleaming diffuse background in almost all
directions. This diffuse X-ray background is considered to be produced
by a combination of several sources, and a clean separation of them
has been a difficult task.

\citet{mcc02} showed that about 40\% of the dimmer all-sky diffuse
emission in the 0.5 -- 1.0 keV band can be attributed to the
extragalactic contribution, which we refer to as the Cosmic X-ray
Background (CXB) in this paper. The remainder of the emission can be
characterized by highly ionized ion emission lines, which are
considered to originate from at least three different sources; the
solar-wind charge-exchange (SWCX)-induced emission from the
Heliosphere (\cite{cox1998}; \cite{cravens2000}; \cite{lal04}), the
thermal emission from the hot gas in local hot bubble (LHB)
(\cite{McCammon:1990p3530}), and another thermal emission from a thick
Galactic hot gaseous disk (\cite{yao07}; \cite{yoshino09};
\cite{yao09}; \cite{hagihara10}). The first two sources are considered
to arise from the local part of the Galaxy and are hard to be
separated from each other. The sum of these emission components can be
approximated by an optically-thin thermal plasma of $kT\sim
0.1\,\mathrm{keV}$ with negligible foreground absorption
(\cite{smith07}; \cite{hen07}; \cite{Galeazzi:2007p3531};
\cite{Kuntz:2008p3532}; \cite{masui09}).

The last component is considered to arise from distant parts of
the Galaxy; mostly beyond the bulk of absorption in the Galactic
disk. The spectrum of the component can be described by thermal plasma
of a temperature of $0.25$~keV. There is so far little constraint
on the distances other than their foreground absorption.
The detection of highly ionized absorption lines of
O\emissiontype{VII}, O\emissiontype{VIII}, and/or Ne\emissiontype{IX}
in spectra of Galactic sources (e.g., \cite{fut04}; \cite{yao05}; \cite{Juett2006di}; \cite{yao08})
indicates that a large fraction of hot gas contributing to the
absorption exists within distance scales of $\sim 10$~kpc. 

\citet{yao07} introduced a joint fitting method of the
high-resolution absorption and emission spectra to constrain not only
the temperature but also the extent and density of the hot gaseous disk.
In the work by Yao \& Wang, the high-resolution absorption spectrum observed with
\emph{Chandra} is jointly fitted with the broadband emission spectrum
from the Rosat All Sky Survey (RASS) of the field adjacent to the
sight line toward the nearby bright active galactic nucleus Mkn
421. This joint fit provides tight constraints on the extent and
density of the hot gas, which is shown to be in a nonisothermal state,
as evidenced by the mean temperature differences inferred from fitting
the absorption and emission spectra independently. The joint fit then
assumes a model of hot gas distribution with the temperature and density
decreasing exponentially with the vertical distance from the Galactic
plane. This exponential model was also used in the joint analysis of
the \emph{Chandra} observed absorption spectrum along the sight line
toward the LMC X--3 with the \emph{Suzaku} observed emission spectra
in the adjacent fields \citep{yao09}. This analysis gives the
estimates of the temperature and density at the Galactic plane as
$(3.6^{+0.8}_{-0.7})\times 10^6\,\mathrm{K}$ and
$(1.4^{+2.0}_{-1.0})\times 10^{-3}\,\mathrm{cm^{-3}}$ and the
exponential scale heights of the hot gas as
$1.4^{+3.8}_{-1.2}\,\mathrm{kpc}$ and
$2.8^{+3.6}_{-1.8}\,\mathrm{kpc}$, respectively.

Furthermore, \citet{hagihara10} applied the same model along the sight
line toward the blazer PKS 2155--304, and estimated the gas
temperature and density at the Galactic plane as
$(2.5^{+0.6}_{-0.3})\times 10^6\,\mathrm{K}$ and
$(1.4^{+0.5}_{-0.4})\times 10^{-3}\,\mathrm{cm^{-3}}$ with the
exponential scale heights of the hot gas as
$5.6^{+7.4}_{-4.2}\,\mathrm{kpc}$ and
$2.3^{+0.9}_{-0.8}\,\mathrm{kpc}$, respectively, which are consistent
with those for the LMC X--3 direction. Both results from \citet{yao09}
and \citet{hagihara10} suggested the presence of a thick hot gaseous
disk with the scale height less than 10~kpc that is consistent with
the Galactic source absorption lines.

The Galactic hot gaseous disk has also been studied for a possibility
of its being a reservoir of missing baryons in the local universe
(\cite{Gupta12}; \cite{Gupta13}; \cite{fang13}; \cite{miller13}).
It is therefore essential to constrain the scale heights of the
hot gas in different parts of the sky individually.
\citet{Gupta12} have argued that the spatial extent of
the hot gas is more than 100~kpc from the joint analysis of
\emph{Chandra} absorption spectra with the emission measure from
literature, but the results are inconsistent with the Galactic
source absorption lines as well as previous studies
(\cite{yao07}; \cite{yao09}; \cite{hagihara10}).
Various issues about the data analysis and interpretation of
\citet{Gupta12} were raised by \citet{wang12}.

In this paper, we revisit the Mkn 421 direction using the \emph{Suzaku}
observed emission spectra to jointly analyze with the absorption spectrum
from \emph{Chandra}. Compared to the \emph{Rosat} detector, the
\emph{Suzaku} CCD has a better sensitivity and a better energy resolution
to separate the hot gaseous disk from the SWCX and LHB ones.

In Section 2, we describe our observations and
data-reduction processes. We discuss our data analysis in Section 3,
and conclude the discussion in Section 4. Throughout the paper, the
statistical errors are quoted at the 90\% confidence level in the
text, as well as in the tables, while the 68\% ($1\sigma$) confidence
errors are adopted in the figures, unless otherwise specified.

\section{Observations and Data Reduction}

\begin{figure}[htbp]
  \begin{center}
    \FigureFile(80mm,50mm){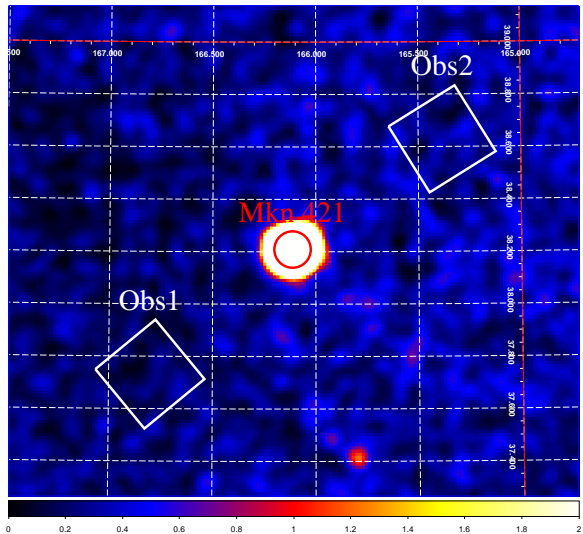}
  \end{center}
  \caption{RASS 0.1 - 2.4 keV band X-ray mapping in the vicinity of
    Mkn 421 (at the center, spread by the \emph{ROSAT} point-spread
    function) and the XIS FOVs of the two presented observations.}
  \label{fig:Observations}
\end{figure}

\subsection{Chandra Observations and Data Reduction}

As one of the brightest X-ray sources, Mkn 421 has been observed by
\emph{Chandra} many times, using the High-Energy Transmission Grating
(HETG) with the Advanced CCD Imaging Spectrometer (ACIS-S) and the
Low-Energy Transmission Grating (LETG) with either the ACIS-S or the
High-Resolution Camera (HRC-S) aboard. The LETG provides $\sim 30$--40~cm$^2$
 effective area at O\emissiontype{VII} and
O\emissiontype{VIII} K$\alpha$ wavelengths with a modest resolution of
$\sim 750$~km s$^{-1}$ (FWHM).

We have used the same absorption spectra as in \citet{yao07}, which
resulted from combining the six long exposures of the existing
observations (Table~\ref{tab:Chandra-Observation-Log}).

\begin{table}
  \begin{center}
    \caption{\emph{Chandra} observation log.}
    \label{tab:Chandra-Observation-Log}
    \begin{tabular}{ccccc}
      \hline
      Obs ID & Obs Date & Grating & Detector & Exposure \\
      & & & & (ks) \\ \hline
      4148 & 2002-10-26 & LETG & ACIS & 96.84 \\
      4149 & 2003-07-01 & LETG & HRC & 99.98 \\
      5171 & 2004-07-03 & LETG & ACIS & 67.15 \\
      5318 & 2004-05-06 & LETG & ACIS & 30.16 \\
      5331 & 2004-07-12 & LETG & ACIS & 69.50 \\
      5332 & 2004-07-14 & LETG & ACIS & 67.06 \\ \hline
    \end{tabular}
  \end{center}
\end{table}

\subsection{Suzaku Observations and Data Reduction}

We observed the emission from the two off-fields of the Mkn 421 sight
line during the AO4 program. The detailed \emph{Suzaku} observation
log of the two fields, Obs1 and Obs2, is listed in
Table~\ref{tab:Suzaku-Observation-Log}.

\begin{table*}
  \begin{center}
    \caption{\emph{Suzaku} observation log.}
    \label{tab:Suzaku-Observation-Log}
    \begin{tabular}{lcc}
      \hline
      & Obs1 & Obs2 \\ \hline
      ($\alpha$, $\delta$) in J2000 ($^\circ$) & (166.80,
      37.73) & (165.38, 38D.63) \\
      ($\ell$, $b$) in Galactic coordinate ($^\circ$) & (180.50,
      65.70) & (179.32, 64.36) \\
      Elongation to Mkn 421 ($^\circ$) & 0.76 & 0.67 \\
      Observation ID & 504086010 & 504087010 \\
      Observation start time (UT) & 2009-11-09T01:34:20 &
      2009-11-11T10:45:08 \\
      Observation end time (UT) & 2009-11-10T19:25:07 &
      2009-11-13T07:37:24 \\
      Exposure time & 75 ks & 86 ks \\
      Exposure after data reduction & 64.3 ks & 74.5 ks \\
      \hline
    \end{tabular}
  \end{center}
\end{table*}

We used the CCD camera, the X-ray Imaging Spectrometer (XIS:
\cite{koyama07}), aboard \emph{Suzaku} (\cite{mitsuda07}) for our
observations.  To minimize stray light from the Mkn 421, the
\emph{Suzaku} XIS fields of view (FOV) of the two observations were
chosen to be at least \timeform{30'} away from the Mkn 421
(Figure~\ref{fig:Observations}). The XIS was set to the normal
clocking mode and the data format was either $3\times 3$ or $5\times
5$. The Spaced-raw Charge Injection (SCI) was on for both the
observations. Processed data version for the observations is
2.4.12.27. In this work, we used the spectra obtained with XIS 1,
which is the backside-illuminated CCD and thus shows superior
sensitivity at photon energies below 1 keV comparing to
frontside-illuminated CCDs, XIS 0 and XIS 3.

We adopted the same data screening as in \citet{hagihara10} to obtain
the good time intervals (GTIs), i.e. excluding exposures when the line
of sight of Suzaku was elevated above the sunlit limb of Earth by less
than $20^\circ$ and exposures with the ``cut-off rigidity'' less than
8 GV.

\subsubsection{Contamination from Solar X-rays}

The obtained spectra may be contaminated by solar X-rays, scattered
off the Earth's atmosphere into the telescope by either Thompson
scattering or fluorescence. As \emph{Suzaku} orbits the Earth with the
fixed pointing, the column density of the atmosphere along the line of
sight, which depends on the elevation of the satellite, varies rapidly
and may affect the intensity of the scattered X-ray intensity. As we
are particularly concerned about O\emissiontype{VII} emission lines
around 0.56 keV, we checked the dependency of the 0.4 to 0.7 keV
photon counting rate on the Oxygen column density of the sunlit
atmosphere in the line of sight using the MSIS atmosphere model (see
\cite{fujimoto07}; \cite{smith07}; \cite{miller08}). The result showed
no correlation, thus there should be no significant neutral Oxygen
emission from the Earth's atmosphere in the filtered data.

\subsubsection{Contamination from the Geocoronal SWCX}

The obtained spectra may also be contaminated by the SWCX induced
emission from the geocorona (\cite{fujimoto07}). The contamination
can be reduced by excluding the exposure time meeting the following
three criteria: the excessive solar-wind proton flux, solar-wind
ion flux, and the small Earth-to-magnetopause (ETM) distance.

\begin{figure}[htbp]
  \begin{center}
    \FigureFile(80mm,50mm){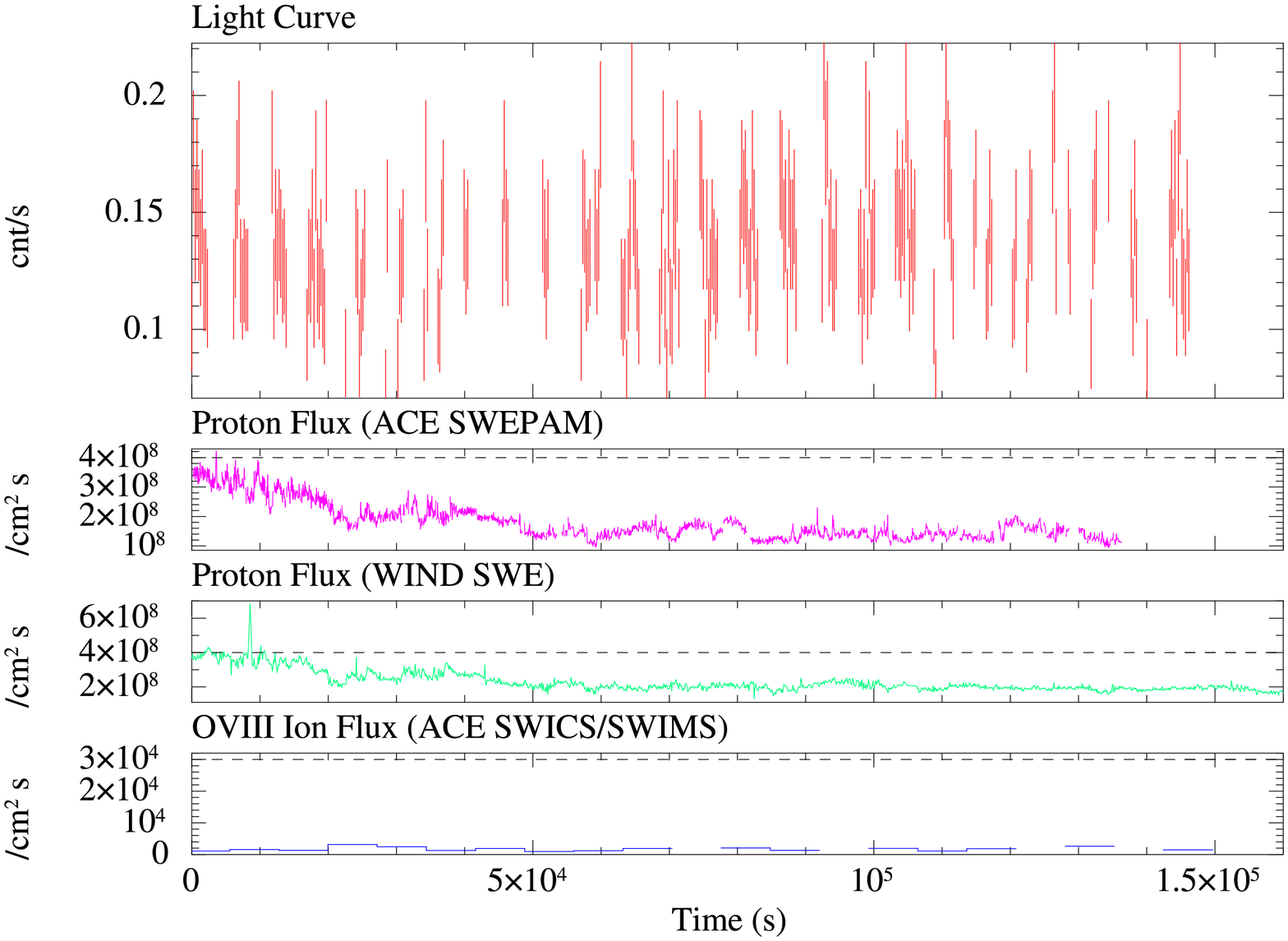} \\
    \FigureFile(80mm,50mm){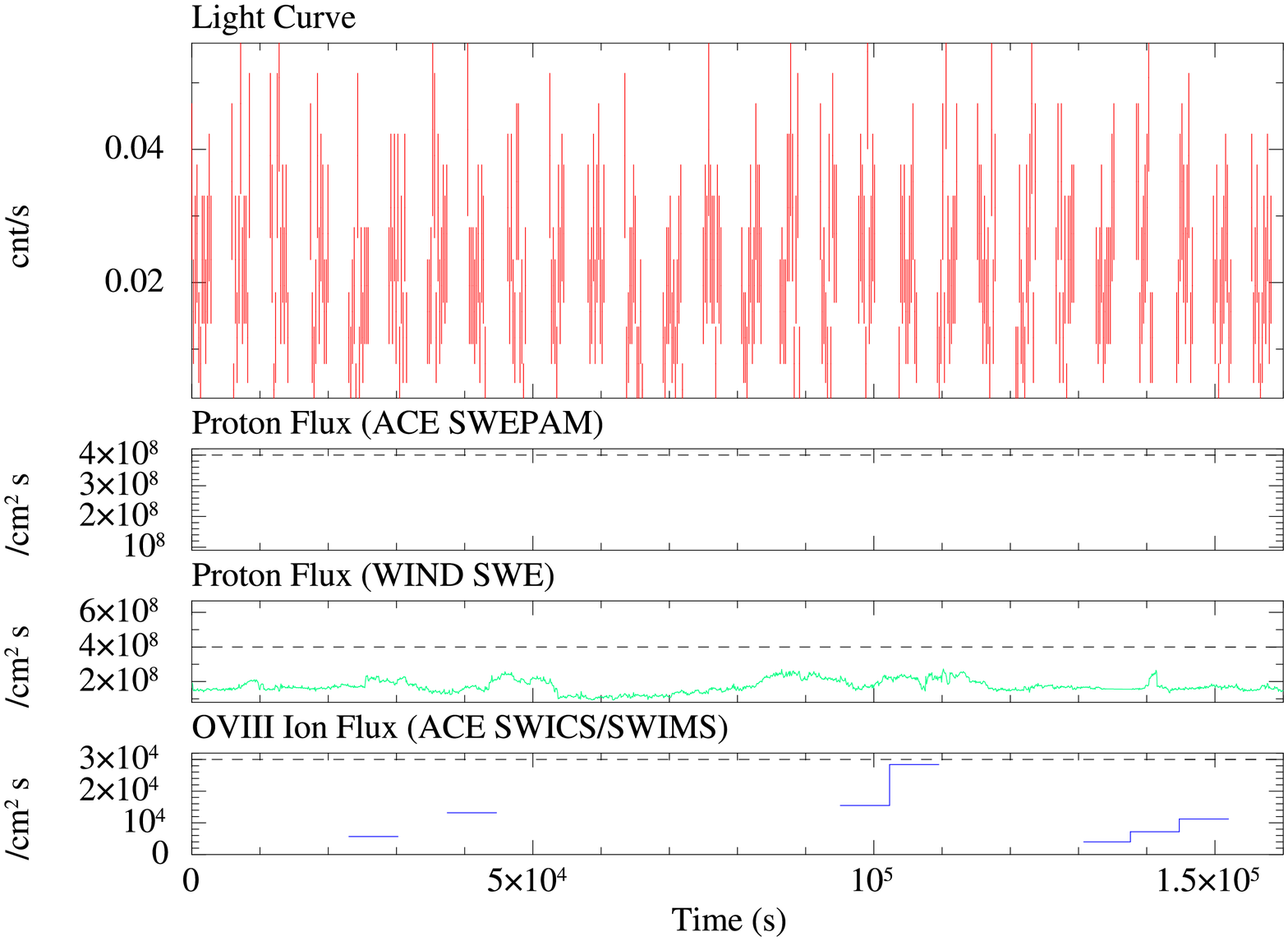}
  \end{center}
  \caption{XIS-BI light curve in 0.5-0.7 keV for Obs1 (top panel) and
    Obs2 (bottom panel), and respective \emph{ACE} SWEPAM (64 sec bin)
    and \emph{WIND} SWE (90 sec bin) observed proton flux, and
    \emph{ACE} SWICS/SWIMS (2 hour average) observed O
    \emissiontype{VIII} ion flux. The origins of both \emph{ACE} and
    \emph{WIND} are set to the beginning of the \emph{Suzaku}
    observations.}
  \label{fig:ProtonFlux}
\end{figure}

\begin{figure}[htbp]
  \begin{center}
    \FigureFile(80mm,50mm){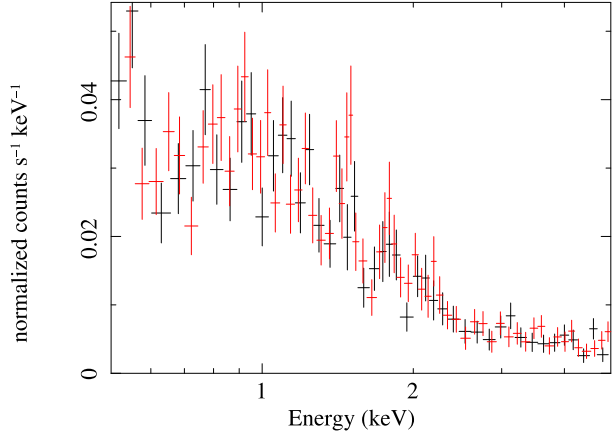}
  \end{center}
  \caption{Energy spectra of Obs1 for the time intervals in the first
    half (black) and the last half (red). There is no significant discrepancy between the two
    spectra.}
  \label{fig:Obs1-Spec-Comparison-for-FH-and-LH}
\end{figure}

For the solar-wind proton flux and ion flux, we used data obtained
with the Solar Wind Electron Proton and Alpha Monitor (SWEPAM), the
Solar Wind Ion Composition Spectrometer (SWICS) and the Solar Wind Ion
Mass Spectrometer (SWIMS) aboard the \emph{Advanced Composition
Explorer (ACE)}, and the Solar Wind Experiment (SWE) aboard the
\emph{WIND}. The light curves of XIS-BI in the energy range of 0.5 to
0.7 keV for Obs1 and Obs2, along with their proton flux and ion flux
are shown in Figure~\ref{fig:ProtonFlux}. We could see short time
periods when the proton flux in the solar wind exceeded the typical
threshold, $4\times 10^8\ \mathrm{cm^{-2}\ s^{-1}}$ (\cite{masui09}),
in the beginning of Obs1 observation, however, we did not remove the
intervals since the O\emissiontype{VIII} ion flux stayed far below the
typical value reported in
\citet{kou06}. Figure~\ref{fig:Obs1-Spec-Comparison-for-FH-and-LH}
shows the spectra of the Obs1 observation for the time intervals in
the first half and the last half. Since we could not see a significant
discrepancy between the two spectra, the exceeded proton flux should
not significantly affect the Obs1 spectrum.  For Obs2, we did not
remove any time intervals either, since both the proton flux and ion
flux are below the threshold throughout the observation.

The last criteria was the ETM distance in the line sight of
\emph{Suzaku} (\cite{fujimoto07}), which should be $> 5R_{\mathrm{E}}$
to avoid the contamination. For both Obs1 and Obs2, the ETM distances
were more than $10R_{\mathrm{E}}$ all the time.

\subsubsection{Point Source Removal}

\begin{figure}
  \begin{center}
      \FigureFile(80mm,50mm){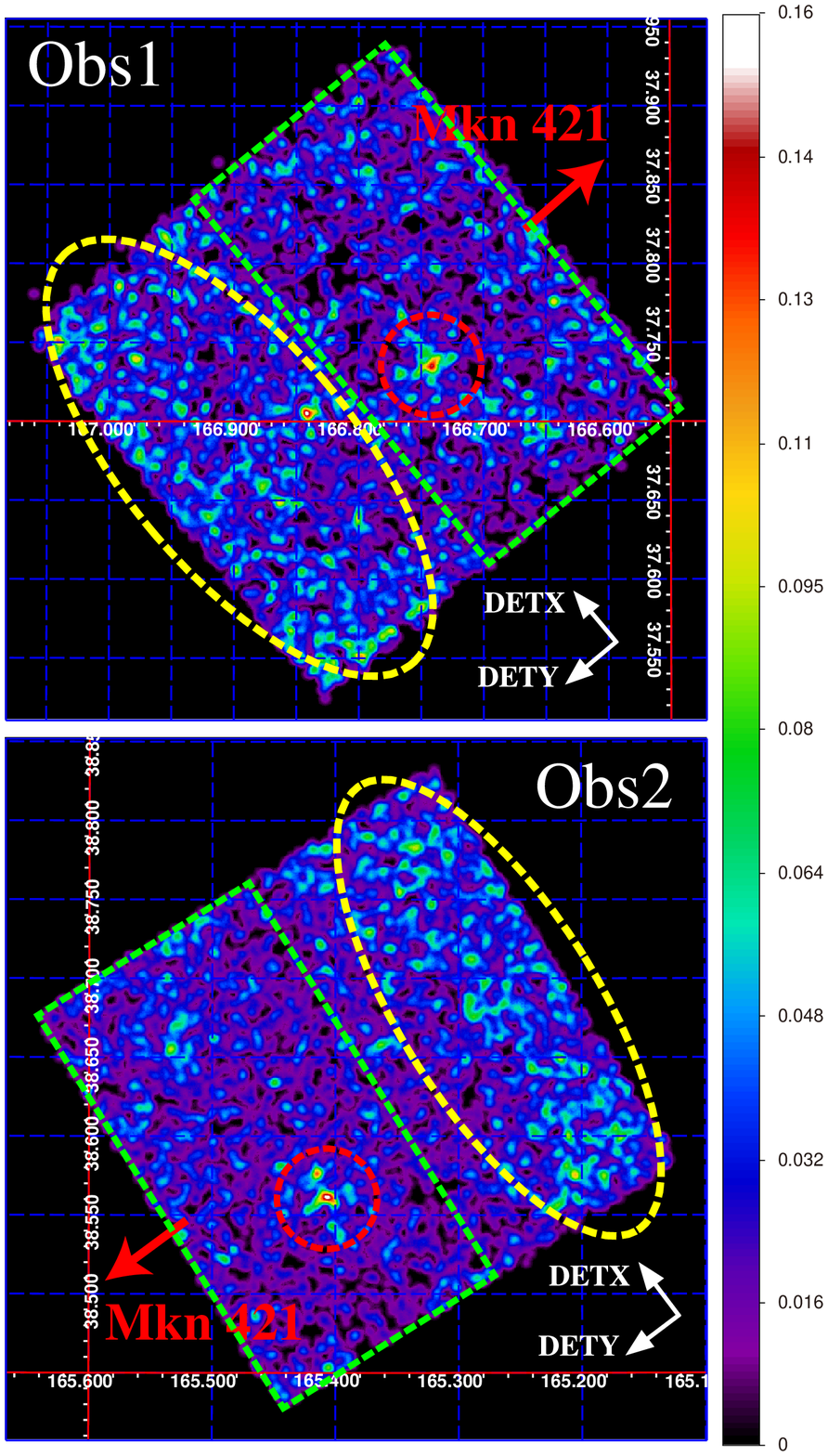}
      \caption{0.5--1.2 keV XIS-BI images of Obs1 (top) and Obs2
        (bottom). Point sources were removed with the radius of
        \timeform{2'} (dashed red circle).  On the opposite sides of
        Mkn 421 (red allow) the diffuse emission is enhanced in both
        observations (dashed yellow circle), which are considered to
        be due to stray light from Mkn 421.  The regions were
        carefully chosen to minimize the contamination from the stray
        light (dashed green box).}
  \label{fig:Stray}
  \end{center}
\end{figure}

We then constructed an image in the 0.5 to 1.2 keV energy range and
detected one point source each in the FOVs of Obs1 and Obs2. We
removed circular regions centered at those point sources with the
radius of \timeform{2'} (Figure~\ref{fig:Stray}). The radius was
determined so that the number of counts from the point sources outside
the circular region is less than 3\% of that from the diffuse X-ray
emission in the energy range.

\subsubsection{Considerable stray light from Mkn 421}

\begin{figure}
  \begin{center}
    \FigureFile(80mm,50mm){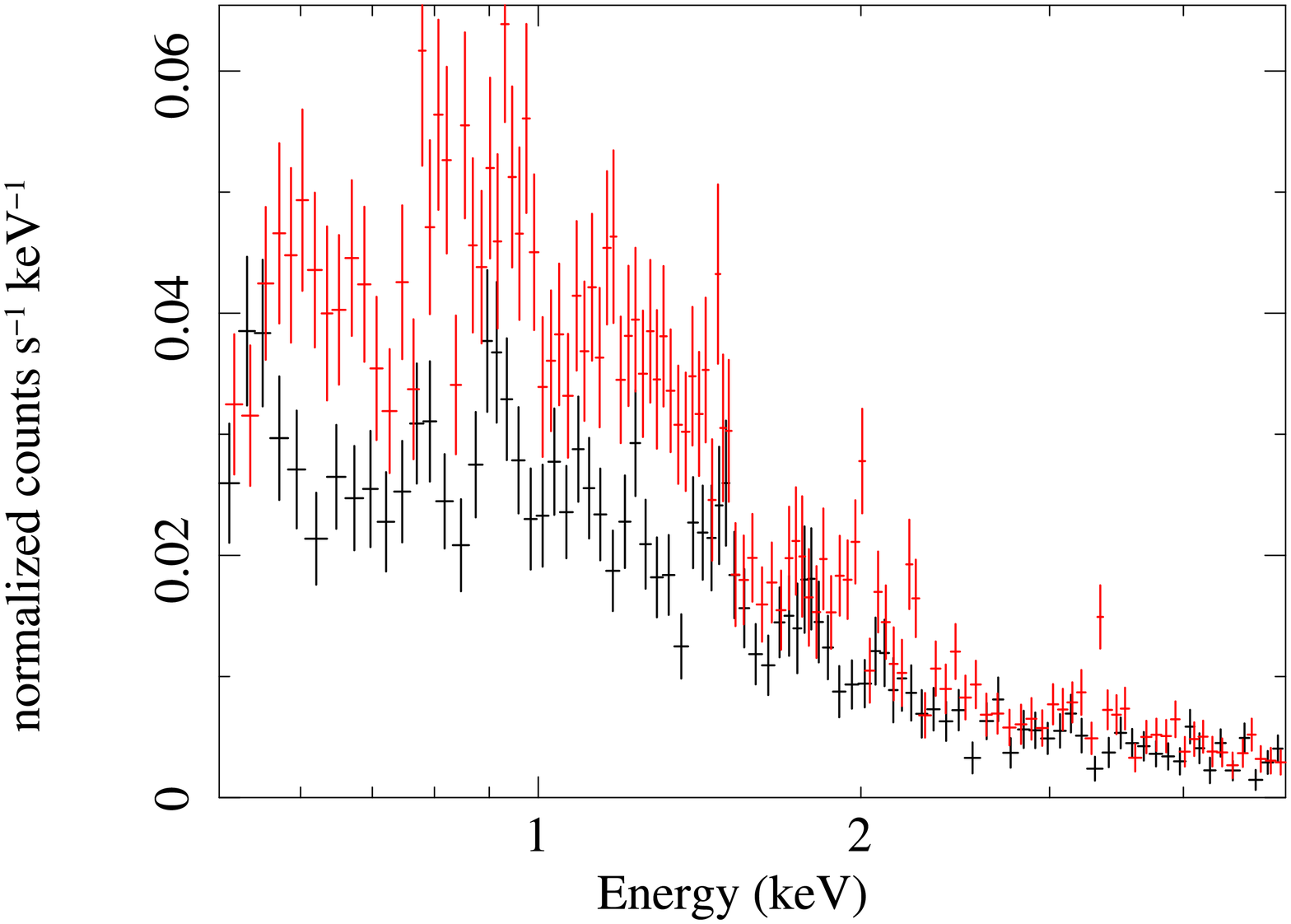} \\
    \FigureFile(80mm,50mm){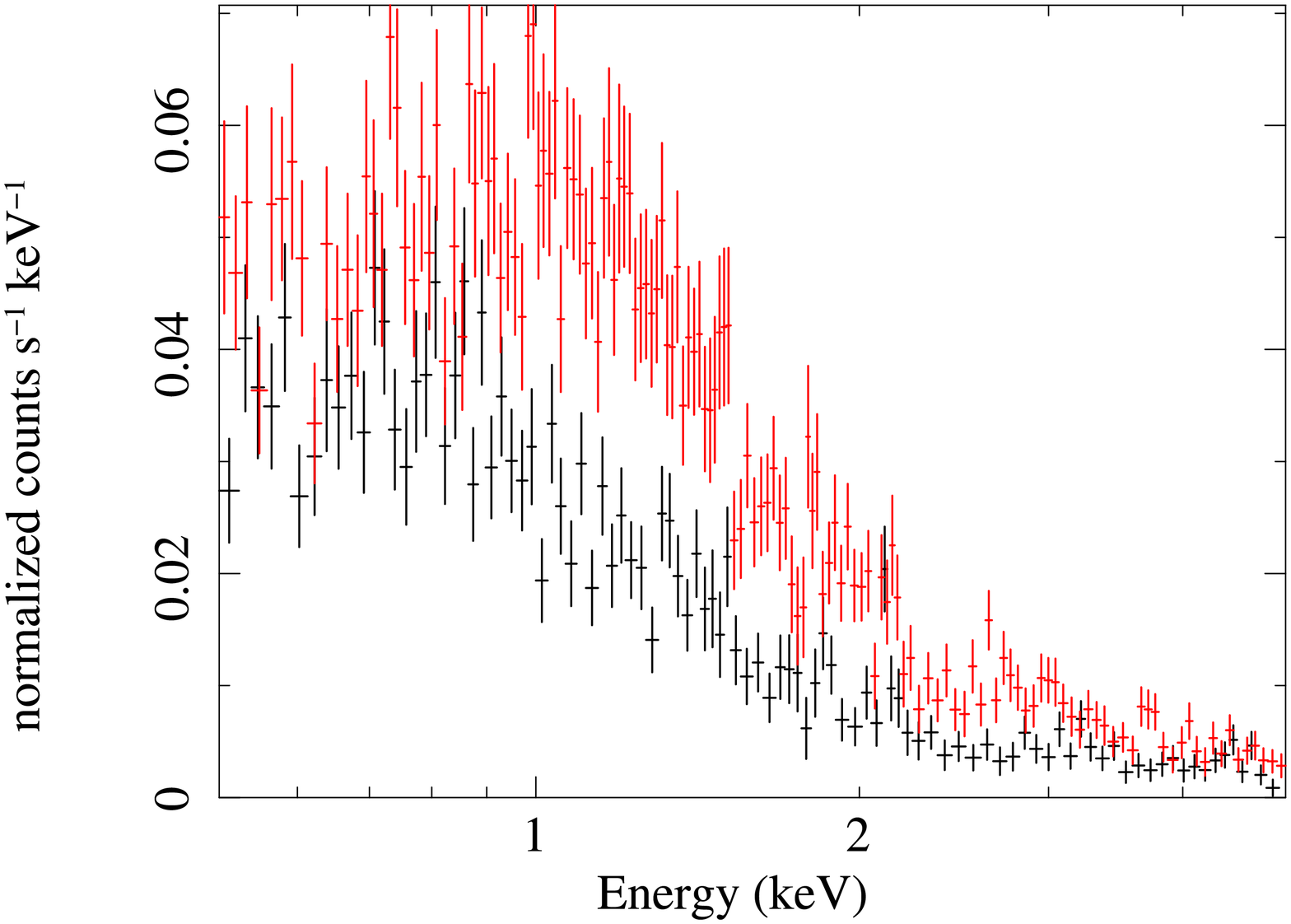}
  \end{center}
  \caption{Energy spectra of the regions, the near side (black) and the far side (red) toward Mkn 421, for
    Obs1 (top) and Obs2 (bottom).}
  \label{fig:Spec-Stray}
\end{figure}

\begin{figure}
  \begin{center}
    \FigureFile(80mm,50mm){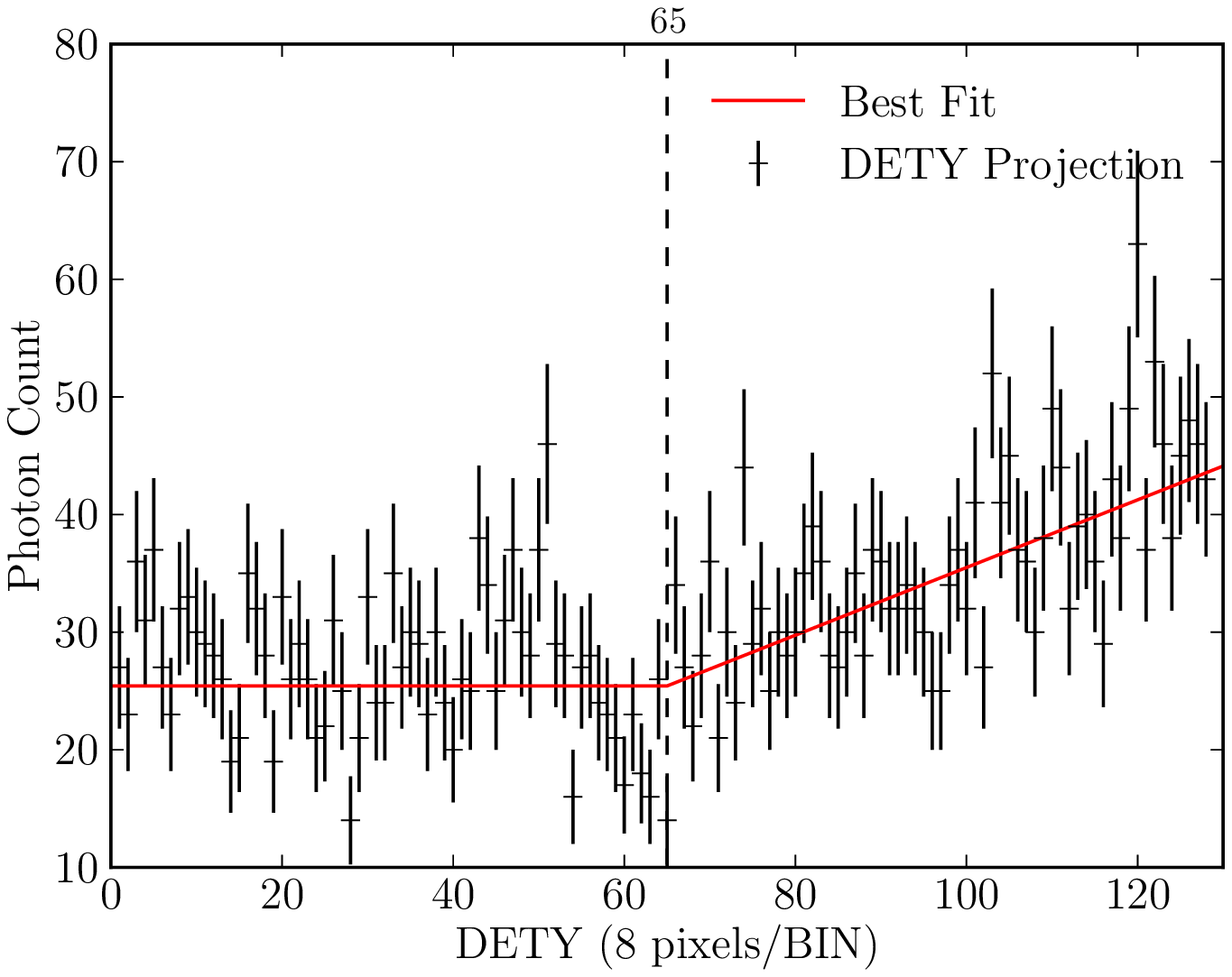} \\
    \FigureFile(80mm,50mm){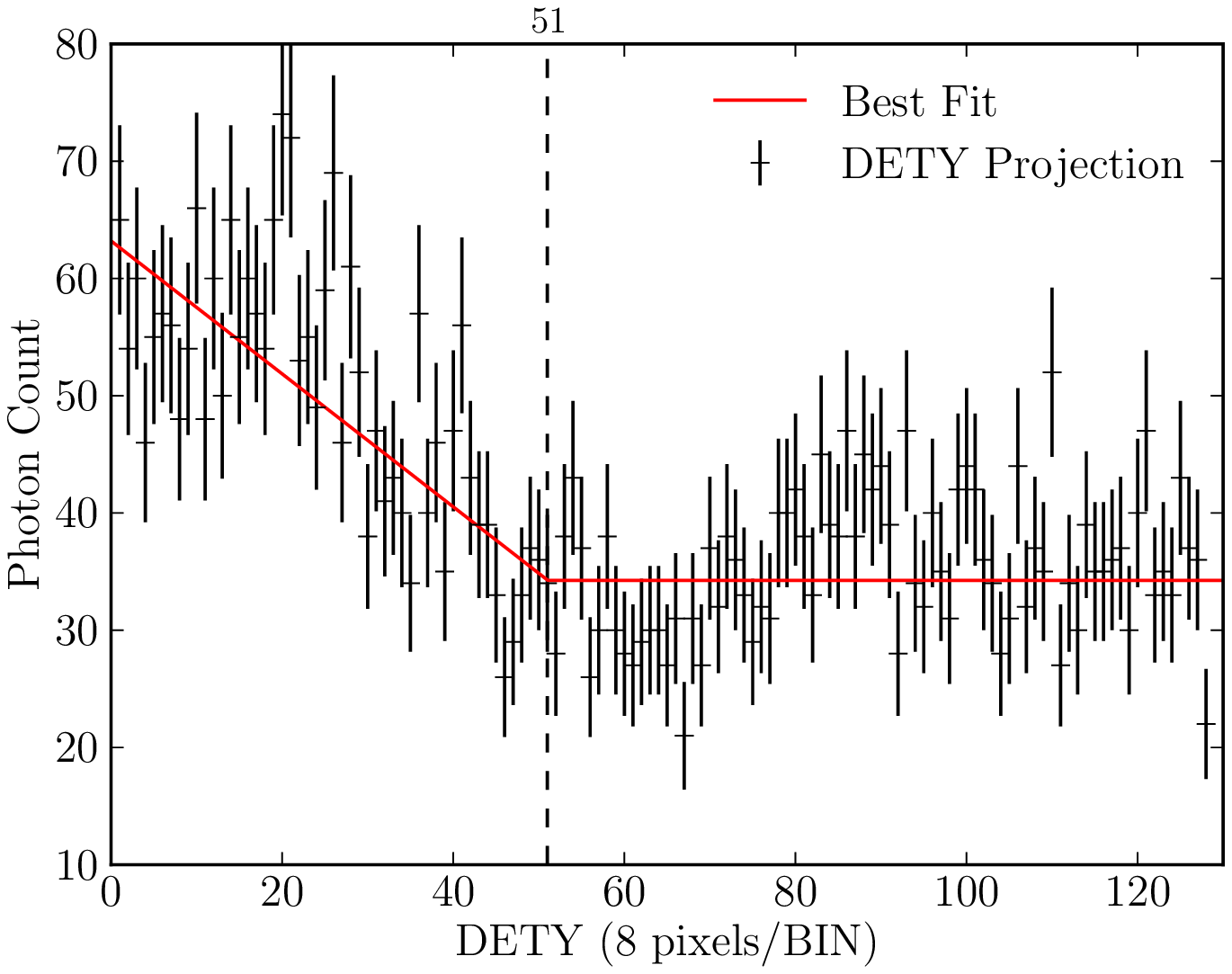}    
    \caption{DETY projected 0.5--1.2 keV photon counts and the
      best-fitted \emph{constant} + \emph{linear} for Obs1 (Top) and
      Obs2 (Bottom).}
  \label{fig:DETY-Projection}
  \end{center}
\end{figure}

In both the \emph{Suzaku} observations, we found a discrepancy in the
brightness between the near side and the far side of the images toward
Mkn 421 (Figure~\ref{fig:Stray}). In both images the far side toward
Mkn 421 is brighter than the other indicating stray light from Mkn 421
likely caused by the secondary refrection at the X-ray telescope
(\cite{Serlemitsos:2007p728,takei239}). Figure~\ref{fig:Spec-Stray} shows the
spectra of the near side and the far side of the images for Obs1 and
Obs2. There are large discrepancies between the two spectra in $<2.0$
keV for both Obs1 and Obs2. To minimize the confusion caused by the
stray light, we used uncontaminated regions for the further
analysis. The regions were determined by the following method. We
first extracted the images for the 0.5 to 1.2 keV range in the DET
coordinates with 8 pixels per bin, then projected the photon count to
the DETY axis. Figure~\ref{fig:DETY-Projection} shows the result of
the projection, along with the best-fitted \emph{constant} +
\emph{linear} given by
\begin{equation}
N_{\gamma}=\left\{
\begin{array}{ll}
  b & (DETY<a)\\
  c\times(DETY-a)+b & (DETY>a)
\end{array}
\right.
\end{equation}
for Obs1 and
\begin{equation}
N_{\gamma}=\left\{
\begin{array}{ll}
  c\times(DETY-a)+b & (DETY<a)\\
  b & (DETY>a)
\end{array}
\right.
\end{equation} for Obs2.
The fitted results are shown in Table~\ref{tab:DETY-Projection}. We
first estimated the maximum (minimum for Obs2) error-allowed
break-point ($a$), 70.4 bins for Obs1 and 47.9 bins for Obs2, and then
calculated the DETY length of the region to maximize the size of the retained
regions (dashed green boxes in Figure~\ref{fig:Stray}) for better
statistics of the extracted emission spectra.

\begin{table}[htbp]
  \begin{center}
    \caption{Fitting results of the DETY projected photon counts in
      the energy range from 0.5 to 1.2 keV with \emph{constant} +
      \emph{linear} for both Obs1 and Obs2 (1$\sigma$ confidence
      level).}
    \label{tab:DETY-Projection}
    {\small
    \begin{tabular}{lccrc}
      \hline
      & Break ($a$) & Const ($b$) & \multicolumn{1}{c}{Slope ($c$)} & $\chi^2$/dof \\ 
      & (bin)\footnotemark[$*$] & (bin)\footnotemark[$*$] & \multicolumn{1}{c}{(bin)\footnotemark[$*$]} \\ \hline
      Obs1 & $65.0\pm 5.4$ & $25.4\pm 0.6$ & $0.29\pm 0.04$ & 165.3/125 \\
      Obs2 & $51.1\pm 3.2$ & $34.2\pm 0.7$ & $-0.57\pm 0.06$ & 151.8/125 \\ \hline 
      \multicolumn{5}{@{}l@{}}{\hbox to 0pt{\parbox{85mm}{\footnotesize
          \par\noindent
          \footnotemark[$*$] 8 pixels per bin.
      }\hss}}
    \end{tabular}
    }
  \end{center}
\end{table}

\subsubsection{Background and response}

ARFs used in this work for \emph{Suzaku} observed spectra were
generated using \texttt{xissimarfgen}, in which source mode is set to
UNIFORM with the radius of \timeform{20'} to take diffuse stray light
effects into account. We then used \texttt{marfrmf} to combine ARFs
with RMFs, which were generated using \texttt{xisrmfgen}, to generate
response files. Non-Xray backgrounds were generated using
\texttt{xisnxbgen} as described in \citet{tawa08}.

\section{Spectral Analysis and Results}

The spectral analysis was performed using XSPEC v12.7.1. We adopted
the solar abundances as given in \citet{and89}.

\subsection{Chandra X-Ray Absorption Spectrum}

\begin{figure}
  \begin{center}
    \FigureFile(80mm,50mm){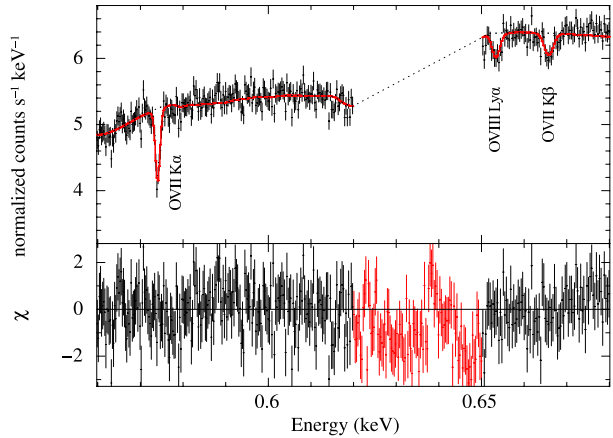}
    \caption{\emph{Chandra} spectrum of Mkn 421 for the 0.55--0.7 keV
      range, fitted with the \emph{wabs}(\emph{power-law} + 3 $\times$
      \emph{Gaussian}) model. The energy range of 0.63 to 0.65~keV was
      ignored due to artificial features seen in the spectrum, but the
      residuals in this range are plotted in red.}
  \label{fig:Spec-Absorption}
  \end{center}
\end{figure}

We first measured the equivalent widths (EWs) of the absorption lines
of the highly ionized oxygen ions, which we have a great interest
in. We therefore fitted the \emph{Chandra}-observed Mkn 421 spectrum
only for the energy range of 0.55 to 0.68~keV using a model, Abs(a),
that consists of a \emph{power-law} and three \emph{gaussian}
functions convolved with absorption by the neutral ISM (\emph{wabs}:
\cite{morrison83}), then obtained the centroids, sigmas and equivalent
widths of O\emissiontype{VII} K$\alpha$, O\emissiontype{VIII}
Ly$\alpha$ and O\emissiontype{VII} K$\beta$ absorption lines as shown
in Figure~\ref{fig:Spec-Absorption}. In the fitting, we ignored the
energy range of 0.63 to 0.65~keV (the residuals in this range are
plotted in Figure~\ref{fig:Spec-Absorption} in red) due to artificial
features (likely caused by the iodine M-edge of HRC) seen in the
spectrum that significantly worsen the goodness of fit. The column
density of neutral hydrogen was fixed to $1.92\times 10^{20}\
\mathrm{atoms\ cm^{-2}}$ determined by the LAB Survey of Galactic
H\emissiontype{I} in this direction \citep{Kalberla}. The fit results
are shown in Table~\ref{tab:Fit-Result-Absorption}. The measured EWs
were consistent with those reported by \citet{wil05}.

\begin{table}
  \begin{center}
    \caption{Spectral fitting of absorption data with
      \emph{wabs(power-law + 3 $\times$ Gaussian)} model.}
    \label{tab:Fit-Result-Absorption}
    \begin{tabular}{cr@{}lr@{}lr@{}l}
      \hline
      & \multicolumn{2}{c}{O\emissiontype{VII} K$\alpha$} & \multicolumn{2}{c}{O\emissiontype{VIII} Ly$\alpha$} & \multicolumn{2}{c}{O\emissiontype{VII} K$\beta$} \\ \hline
      Centroid (eV) & $574.0$ & $^{+0.1}_{-0.1}$ & $653.4$ & $^{+0.4}_{-0.3}$ & $665.8$ & $^{+0.4}_{-0.4}$ \\
      Sigma (eV) & $0.18$ & $^{+0.14}_{-0.18}$ & $0.65$ & $^{+0.54}_{-0.39}$ & $0.84$ & $^{+0.68}_{-0.56}$ \\
      EW (eV) & $0.295$ & $^{+0.022}_{-0.024}$ & $0.138$ & $^{+0.028}_{-0.031}$ & $0.145$ & $^{+0.033}_{-0.034}$ \\ \hline
    \end{tabular}
  \end{center}
\end{table}

We then replaced the \emph{gaussian} functions by \emph{absem}
functions. The \emph{absem}, which is a revision of the \emph{absline}
function of \citet{yao05}, can be used to jointly fit the absorption
and the emission spectra (see \cite{yao07} and \cite{yao09} for a
detailed description).

Assuming a gas with a uniform density and a single temperature, we
first performed a joint analysis of O\emissiontype{VII} K$\alpha$ and
O\emissiontype{VII} K$\beta$ to constrain the O\emissiontype{VII}
column density and the velocity dispersion $v_b$. With the constrained
$v_b$, adding the O\emissiontype{VIII} K$\alpha$ line in the analysis
also yields the column density of O\emissiontype{VIII}. We call this
model as Abs(b). Since the ratio of column densities of
O\emissiontype{VII} and O\emissiontype{VIII} is sensitive to the gas
temperature, a joint analysis named Abs(c) of the O\emissiontype{VII}
and O\emissiontype{VIII} lines constrains the gas temperature. Assuming
the solar abundance for oxygen, and given the constrained gas
temperature, the O\emissiontype{VII} (or O\emissiontype{VIII}) column
density can be converted to the corresponding hot-phase hydrogen
column density. This final model is Abs(d). The results of each steps
are listed in Table~\ref{tab:Abs-Fitting-Result}. The constrained
O\emissiontype{VII} column density, $(8.1^{+3.9}_{-3.1})\times
10^{15}\ \mathrm{cm^{-2}}$, is comparable to typical values $\sim
10^{16}\ \mathrm{cm^{-2}}$ obtained from AGN observations given in two
systematic studies (\cite{fang06}; \cite{bre07}).

\begin{table*}
  \begin{center}
    \caption{Spectral fitting result of absorption data with
      \emph{wabs}(\emph{power-law})$\times$\emph{absem}$\times$\emph{absem}$\times$\emph{absem}
      model.}
    \label{tab:Abs-Fitting-Result}    
    \begin{tabular}{lccccccc}
      \hline
      & Galactic & \multicolumn{5}{c}{Hot ISM} \\ \cline{3-7}
      & $\Gamma$ & $v_b$ & $\log T$ & $\log N_{\mathrm{O\emissiontype{VII}}}$ & $\log N_{\mathrm{O\emissiontype{VIII}}}$ & $\log N_{\mathrm{H}}$ & $\chi^2$/dof \\
      & & ($\mathrm{km\ s^{-1}}$) & (K) & ($\mathrm{cm^{-2}}$) & ($\mathrm{cm^{-2}}$) & ($\mathrm{cm^{-2}}$) \\ \hline
      Abs(b) & $2.24^{+0.04}_{-0.04}$ & $60^{+45}_{-13}$ & \ldots & $15.98^{+0.17}_{-0.20}$ & $15.22^{+0.14}_{-0.16}$ & \ldots & 284.81/294 \\
      Abs(c) & $2.24^{+0.04}_{-0.04}$ & $62^{+43}_{-16}$ & $6.18^{+0.03}_{-0.04}$ & $15.96^{+0.19}_{-0.17}$ & \ldots & \ldots & 284.84/294 \\
      Abs(d) & $2.24^{+0.04}_{-0.04}$ & $60^{+45}_{-14}$ & $6.17^{+0.03}_{-0.03}$ & \ldots & \ldots & $19.13^{+0.16}_{-0.18}$ & 284.81/294 \\ \hline
    \end{tabular}
  \end{center}
\end{table*}

\subsection{Suzaku X-Ray Emission Spectra}

\begin{figure}
  \begin{center}
    \FigureFile(80mm,50mm){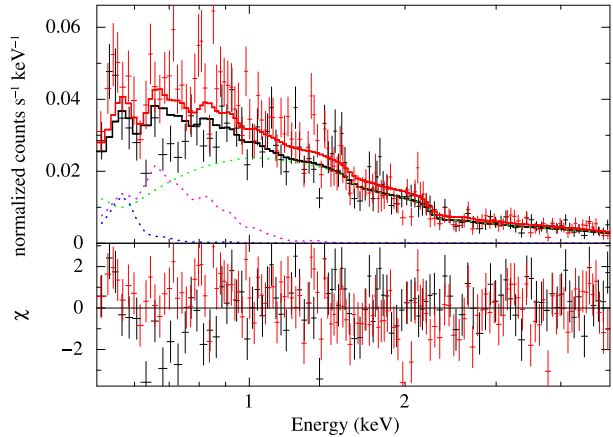}
    \caption{\emph{Suzaku} spectrum of Obs1 (black) and Obs2
      (red), and the best-fitted model convolved with the instrument
      response function. The residuals of the fit are shown in the
      lower panel. The model function, denoted as Emission(A),
      consists of three spectral components; a \emph{power-law} for
      the CXB (green), a \emph{Mekal} for the hot ISM (magenta), and
      another \emph{Mekal} for the SWCX+LHB (blue). The components of
      Obs2 are intentionally omitted in the figure for clarity.}
  \label{fig:Spec-Emission}
  \end{center}
\end{figure}

We first individually fitted the two \emph{Suzaku} emission spectra
extracted for the background emission in the Mkn 421 field. The model
function of the fits consists of an absorbed \emph{power-law} for the
CXB, an absorbed \emph{Mekal} for the hot ISM, and an unabsorbed
\emph{Mekal} for the SWCX and LHB contributions. The temperature and
normalization of the SWCX and LHB component are fixed initially so
that it produces a 2.0 LU O\emissiontype{VII} line intensity, which is
a typical intensity from Heliospheric SWCX and LHB
(\cite{yoshino09}). The systematic error of this step is discussed in
Section \ref{sec:uncertainty-due-swcx}. The fitted results are shown
in Table~\ref{tab:Emi-Fit-Results} as Obs1 and Obs2(a). We then
jointly fitted the Obs1 and Obs2 spectra with the same model
(Figure~\ref{fig:Spec-Emission}); all parameters, except for the
normalizations of the CXB component, were linked between the Obs1 and
Obs2. The results are shown in the same table as Emission(A).

The $\chi^2$ of this joint fit is worse than those of fits to the Obs1
and Obs2 spectra, individually, apparently due to the excess seen in
the Obs2 spectrum in the energy range from 0.6 to 1.0 keV. Similar
excesses have been seen in the \emph{Suzaku} spectra of the diffuse
emission. \citet{yoshino09} reported that the excess can be explained
by either over abundances of some metal elements or an additional
higher temperature emission component with solar abundances. The
latter explanation is considered to be plausible due to the putative
presence of blobs of high ($> 0.22$ keV) temperature hot gas in the
fields \citep{yoshino09}.

\begin{figure}
  \begin{center}
    \FigureFile(80mm,50mm){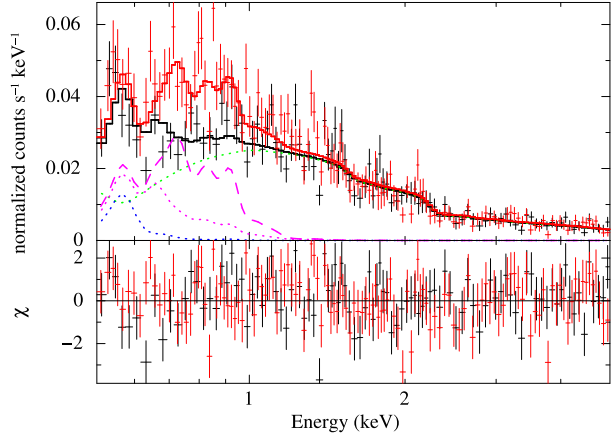}
    \caption{\emph{Suzaku} spectrum of Obs1 (black) and Obs2
      (red), and the best-fitted model and its components. The model
      function, denoted as Emission(B), consists of three spectral
      components; a \emph{power-law} for the CXB (green), a
      \emph{Mekal} for the hot ISM (dotted-magenta for Obs1 and
      dashed-magenta for Obs2), and another \emph{Mekal} for the
      SWCX+LHB (blue). For Obs2, the hot-ISM component is only shown
      in the figure for clarity.}
  \label{fig:Spec-Emission-Ab}
  \end{center}
\end{figure}

\begin{figure}
  \begin{center}
    \FigureFile(80mm,50mm){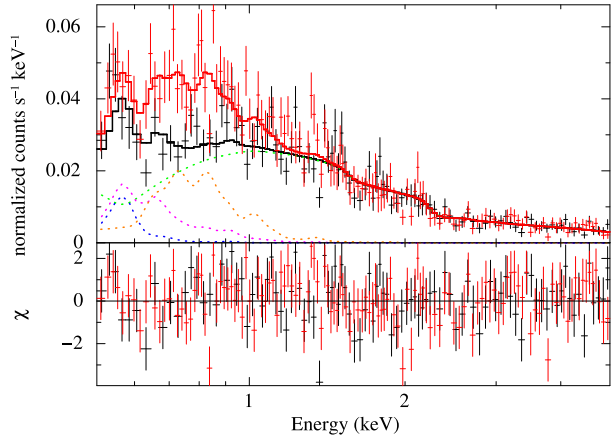}
    \caption{\emph{Suzaku} spectrum of Obs1 (black) and Obs2
      (red), and the best-fitted model and its components. The model
      function consists of four spectral components; a
      \emph{power-law} for the CXB (green), a \emph{Mekal} for the hot
      ISM (magenta), a \emph{Mekal} for the SWCX+LHB (blue), and a
      \emph{Mekal} for a LTC emission (orange). The components of Obs2
      except the LTC emission are intentionally omitted in the figure
      for clarity.}
  \label{fig:Spec-Emission-2T}
  \end{center}
\end{figure}

So we carried out additional fits for Obs2; a fit with an absorbed
\emph{power-law} for the CXB, an absorbed \emph{Mekal} for the hot
ISM, of which Ne and Fe abundances were allowed to vary, and an
unabsorbed \emph{Mekal} for the SWCX and LHB
(Figure~\ref{fig:Spec-Emission-Ab}), and a fit with an absorbed
\emph{power-law} for the CXB, an absorbed \emph{Mekal} for the hot
ISM, an unabsorbed \emph{Mekal} for the SWCX and LHB, and one more
\emph{Mekal} for a local thermal component (hereafter LTC)
(Figure~\ref{fig:Spec-Emission-2T}). The fitted results of those three
are listed in Table~\ref{tab:Emi-Fit-Results} as Obs2(b) and Obs2(c),
respectively. The $\chi^2$ values were then improved in both fits,
however, the temperature of hot ISM for Obs2(b) is still slightly
higher that that of Obs1 while that of Obs2(c) is almost comparable.
We then combined them with the Obs1 spectrum. The results are listed
in the same table as Emission(B) and Emission(C) that correspond to
joint fits of Obs1 with Obs2(b) and Obs2(c) respectively.
In both fits, the $\chi^2$ values were also improved compared with
that of Emission(a).

% $^{}_{}$
\begin{table*}
  \begin{center}
    \caption{Spectral fitting results of the emission data.}
    \label{tab:Emi-Fit-Results}
    {\small
    \begin{tabular*}{180mm}{@{\extracolsep{\fill}}lcccccccccc}
      \hline
      & CXB & \multicolumn{2}{c}{LHB+SWCX} & \multicolumn{4}{c}{Hot ISM} & \multicolumn{2}{c}{LTC} & $\chi^2$/dof \\ \cline{5-8}
      & Norm\footnotemark[$\dagger$] & $\log T$ (K) & Norm\footnotemark[$\ddagger$] & $\log T$ (K) & Norm\footnotemark[$\ddagger$] & Ne/O & Fe/O & $\log T$ (K) & Norm\footnotemark[$\ddagger$] \\ \hline
%%% Single PL (Xspec 12/fits-pasj8) %%%
      Obs1 & $12.2^{+0.5}_{-0.5}$ & 6.06 (fix) & 6.6 (fix) & $6.35^{+0.07}_{-0.05}$ & $2.1^{+0.6}_{-0.6}$ & 1 (fix) & 1 (fix) & $\cdots$ & $\cdots$ & 123.77/99 \\ \hline
      Obs2(a) & $10.9^{+0.5}_{-0.5}$ & 6.06 (fix) & 6.6 (fix) & $6.46^{+0.03}_{-0.03}$ & $3.1^{+0.4}_{-0.4}$ & 1 (fix) & 1 (fix) & $\cdots$ & $\cdots$ & 144.44/126 \\ \hline
      Obs2(b) & $10.8^{+0.5}_{-0.5}$ & 6.06 (fix) & 6.6 (fix) & $6.44^{+0.06}_{-0.10}$ & $3.0^{+0.5}_{-0.5}$ & $2.1^{+2.6}_{-1.0}$ & $1.2^{+2.6}_{-0.6}$ & $\cdots$ & $\cdots$ & 140.90/124 \\ \hline
      Obs2(c) & $10.6^{+0.5}_{-0.5}$ & 6.06 (fix) & 6.6 (fix) & $6.37^{+0.07}_{-0.07}$ & $3.1^{+0.6}_{-0.6}$ & 1 (fix) & 1 (fix) & $6.83^{+0.11}_{-0.30}$ & $0.7^{+0.4}_{-0.4}$ & 135.25/124 \\ \hline
      \multicolumn{2}{@{\extracolsep{\fill}}l}{Emission(A)} & & & & & & & & & 313.32/225 \\ \cline{1-1}
      \ Obs1 & $11.4^{+0.5}_{-0.5}$ & 6.06 (fix) & 6.6 (fix) & $6.44^{+0.03}_{-0.02}$ & $2.4^{+0.3}_{-0.3}$ & 1 (fix) & 1 (fix) & $\cdots$ & $\cdots$ \\
      \ Obs2(a) & $11.5^{+0.4}_{-0.4}$ & $\uparrow$ & $\uparrow$ & $\uparrow$ & $\uparrow$ & $\uparrow$ & $\uparrow$ & $\cdots$ & $\cdots$ \\ \hline
      \multicolumn{2}{@{\extracolsep{\fill}}l}{Emission(B)} & & & & & & & & & 271.83/223 \\ \cline{1-1}
      \ Obs1 & $12.0^{+0.5}_{-0.5}$ & 6.06 (fix) & 6.6 (fix) & $6.35^{+0.04}_{-0.03}$ & $2.6^{+0.4}_{-0.4}$ & 1 (fix) & 1 (fix) & $\cdots$ & $\cdots$ \\
      \ Obs2(b) & $11.0^{+0.5}_{-0.5}$ & $\uparrow$ & $\uparrow$ & $\uparrow$ & $\uparrow$ & $4.8^{+2.3}_{-1.7}$ & $4.2^{+3.0}_{-1.4}$ & $\cdots$ & $\cdots$ \\ \hline
      \multicolumn{2}{@{\extracolsep{\fill}}l}{Emission(C)} & & & & & & & & & 261.00/223 \\ \cline{1-1}
      \ Obs1 & $12.2^{+0.5}_{-0.5}$ & 6.06 (fix) & 6.6 (fix) & $6.33^{+0.04}_{-0.06}$ & $2.2^{+0.6}_{-0.5}$ & 1 (fix) & 1 (fix) & $\cdots$ & $\cdots$ \\
      \ Obs2(c) & $10.8^{+0.5}_{-0.5}$ & $\uparrow$ & $\uparrow$ & $\uparrow$ & $\uparrow$ & $\uparrow$ & $\uparrow$ & $6.59^{+0.22}_{-0.07}$ & $1.4^{+0.5}_{-0.6}$ \\ \hline
      \multicolumn{11}{@{}l@{}}{\hbox to 0pt{\parbox{180mm}{\footnotesize
%          \par\noindent
%          \footnotemark[$*$] Hard limit.
          \par\noindent
          \footnotemark[$\dagger$] In units of $\mathrm{photons\,keV^{-1}\,cm^{-2}\,s^{-1}\,str^{-1}@1keV}$.
          \par\noindent
          \footnotemark[$\ddagger$] In units of $10^{-3}\,\mathrm{pc\,cm^{-6}}$. The emission measure of LHB+SWCX is fixed to $0.0066\,\mathrm{pc\,cm^{-6}}$ which corresponds to 2.0 LU of O\emissiontype{VII} K$\alpha$ emission.
      }\hss}}
    \end{tabular*}
    }
  \end{center}
\end{table*}

Finally we reset the oxygen abundance of the \emph{Mekal} components
that represent the SWCX/LHB and hot ISM to zero, and added two delta
functions at 0.574 keV and 0.654 keV to evaluate the line intensities
of the O\emissiontype{VII} and O\emissiontype{VIII} emissions. Results
are shown in Table~\ref{tab:Emi-Fit-SB-Results}.

\begin{table*}
  \begin{center}
    \caption{O\emissiontype{VII} and O\emissiontype{VIII} line intensities.}
    \label{tab:Emi-Fit-SB-Results}
    {\small
    \begin{tabular*}{180mm}{@{\extracolsep{\fill}}lccccccccc}
      \hline
      & CXB & \multicolumn{2}{c}{Hot ISM} & \multicolumn{2}{c}{LTC} & O\emissiontype{VII} & O\emissiontype{VIII} & Flux\footnotemark[$\|$] & $\chi^2$/dof \\
      & Norm\footnotemark[$\dagger$] & $\log T$ (K) & Norm\footnotemark[$\ddagger$] & $\log T$ (K) &Norm\footnotemark[$\ddagger$] & SB\footnotemark[$\S$] & SB\footnotemark[$\S$] \\ \hline
%%% Single PL (Xspec 12/fits-pasj8) %%%
      Obs1 & $12.0^{+0.5}_{-0.5}$ & 6.35 (fix) & $4.8^{+1.5}_{-1.5}$ & $\cdots$ & $\cdots$ & $4.8^{+1.1}_{-1.1}$ & $0.6 (<1.1)$ & 2.38 & 118.99/96 \\ \hline
      Obs2(a) & $10.9^{+0.5}_{-0.5}$ & 6.46 (fix) & $ $$4.0^{+0.5}_{-0.5}$ & $\cdots$ & $\cdots$ & $5.8^{+1.0}_{-1.0}$ & $2.0^{+0.6}_{-0.6}$ & 2.86 & 144.48/125 \\ \hline
      Obs2(b) & $10.8^{+0.5}_{-0.5}$ & 6.44 (fix) & $3.6^{+0.5}_{-0.5}$ & $\cdots$ & $\cdots$ & $5.8^{+1.0}_{-1.0}$ & $2.0^{+0.6}_{-0.6}$ & 2.85 & 145.60/125 \\ \hline
      Obs2(c) & $10.7^{+0.5}_{-0.5}$ & 6.37 (fix) & $6.2^{+2.8}_{-2.8}$ & 6.83 (fix) & $0.4^{+0.4}_{-0.4}$ & $5.5^{+1.1}_{-1.1}$ & $1.9^{+0.6}_{-0.6}$ & 2.88 & 140.26/123 \\ \hline
      \multicolumn{2}{@{\extracolsep{\fill}}l}{Emission(A)} & & & & & & & & 310.37/224 \\ \cline{1-1}
      \ Obs1 & $11.3^{+0.5}_{-0.5}$ & 6.44 (fix) & $3.2^{+0.4}_{-0.4}$ & $\cdots$ & $\cdots$ & $5.4^{+0.8}_{-0.8}$ & $1.3^{+0.4}_{-0.4}$ & 2.60 \\
      \ Obs2(a) & $11.4^{+0.4}_{-0.4}$ & $\uparrow$ & $\uparrow$ & $\cdots$ & $\cdots$ & $\uparrow$ & $\uparrow$ & 2.60 \\ \hline
      \multicolumn{2}{@{\extracolsep{\fill}}l}{Emission(B)} & & & & & & & & 284.38/224 \\ \cline{1-1}
      \ Obs1 & $12.1^{+0.5}_{-0.5}$ & 6.35 (fix) & $3.0^{+0.4}_{-0.4}$ & $\cdots$ & $\cdots$ & $5.5^{+0.8}_{-0.8}$ & $1.3^{+0.4}_{-0.4}$ & 2.40 \\
      \ Obs2(b) & $11.0^{+0.5}_{-0.4}$ & $\uparrow$ & $\uparrow$ & $\cdots$ & $\cdots$ & $\uparrow$ & $\uparrow$ & 2.74 \\ \hline
      \multicolumn{2}{@{\extracolsep{\fill}}l}{Emission(C)} & & & & & & & & 261.07/223 \\ \cline{1-1}
      \ Obs1 & $12.0^{+0.5}_{-0.5}$ & 6.33 (fix) & $5.6^{+1.6}_{-1.6}$ & $\cdots$ & $\cdots$ & $4.9^{+0.8}_{-0.8}$ & $0.8^{+0.4}_{-0.4}$ & 2.42 \\
      \ Obs2(c) & $10.7^{+0.5}_{-0.5}$ & $\uparrow$ & $\uparrow$ & 6.59 (fix) & $1.2^{+0.3}_{-0.3}$ & $\uparrow$ & $\uparrow$ & 2.85 \\ \hline
      \multicolumn{9}{@{}l@{}}{\hbox to 0pt{\parbox{180mm}{\footnotesize
          \par\noindent
          \footnotemark[$\dagger$] In units of $\mathrm{photons\,keV^{-1}\,cm^{-2}\,s^{-1}\,str^{-1}@1keV}$.
          \par\noindent
          \footnotemark[$\ddagger$] In units of $10^{-3}\,\mathrm{pc\,cm^{-6}}$.
          \par\noindent
          \footnotemark[$\S$] The surface brightness in units of $\mathrm{LU} = \mathrm{photons\,cm^{-2}\,s^{-1}\,str^{-1}}$.
          \par\noindent
          \footnotemark[$\|$] 0.5--1.2 keV range, in units of $10^{-12}\,\mathrm{erg\,cm^{-2}\,s^{-1}}$.
      }\hss}}
    \end{tabular*}
    }
  \end{center}
\end{table*}

By comparing the obtained line intensities of O\emissiontype{VII} and
O\emissiontype{VIII} for Obs1 and Obs2 individual fits to their
joint fits, we could see that O\emissiontype{VII} and
O\emissiontype{VIII} line intensities from Emission(C) are consistent
with that from the Obs1 individual fit. For the
further analysis, we shall therefore employ the model used in
Emission(C).

\subsection{Combined Analysis}

If we assume that the hot ISM observed with \emph{Chandra} and the hot
ISM observed with \emph{Suzaku} originate from the same hot gas, i.e.,
the hot ISM is uniform across the angular extent of the observed fields,
the column density of absorption spectrum and the emission measure of
emission spectra linked with common parameters should estimate the plasma
scale and temperature.

If the ISM were an isothermal plasma with a uniform density in the
solar abundance, the column density of hydrogen ($CD$) and the
emission measure ($EM$) are then given as \begin{eqnarray}
  CD & = & nL\\
  EM & = & (1+2A_{\mathrm{He}})n^{2}L\end{eqnarray}
where $n$ is the hydrogen number density, $A_{\mathrm{He}}$
is the abundance ratio of He to H, and $L$ is the extent of the plasma
along the line of sight. Once $CD$ and $EM$ are determined from
observations, we can derive $n_{\mathrm{H}}$ and $L$ individually.

However, the temperature of the hot ISM obtained from the emission
analysis is higher than that obtained from the absorption analysis. To
account for the difference, \citet{yao09} introduced an exponential
disk model. In this model, we assume that the hydrogen number density
and temperature can be expressed as \begin{equation}
  n=n_{0}e^{-z/(h_{n}\xi)}\qquad\mathrm{and}\qquad
  T=T_{0}e^{-z/(h_{T}\xi)}\label{eq:Exponential-Model}\end{equation}
where $z$ is the vertical distance from the Galactic plane, $n_{0}$
and $T_{0}$ are the density and temperature at the plane, $h_{n}$ and
$h_{T}$ are the scale heights of density and temperature, and $\xi$ is
the volume filling factor, which is assumed to be 1 in this work.

We can now corollary derive \begin{equation}
  n=n_{0}\left(T/T_{0}\right)^{\gamma}\end{equation} where
$\gamma\equiv h_{T}/h_{n}$. The differential hydrogen column density
distribution can then be deduced as \begin{equation}
  dN_{\mathrm{H}}=ndL=\frac{N_{\mathrm{H}}\gamma}{T_{0}}\left(\frac{T}{T_{0}}\right)^{\gamma-1}dT.\end{equation}

Meanwhile, the line intensity $I$ can be expressed as \begin{eqnarray}
  I=\frac{1}{4\pi}\int_{T_{\mathrm{min}}}^{T_{0}}\Lambda(T)\frac{dEM}{dT}dT,\end{eqnarray}
where $\Lambda(T)$ is the emissivity of the hot gas, and the differential EM is\begin{equation}
  dEM=n_{e}n_{\mathrm{H}}dL=\frac{1.2{N_{\mathrm{H}}}^{2}\gamma}{T_{0}L}\left(\frac{T}{T_{0}}\right)^{2\gamma-1}dT,\end{equation}
where $L=h_{n}\xi/\sin b$.

To perform a joint analysis using the \emph{Chandra} absorption
spectrum and the \emph{Suzaku} emission spectra, we first replaced the
\emph{Mekal} functions used in the emission analysis by \emph{vabmkl}
functions (\cite{yao09}) that directly take $T_{0}$, $N_{\mathrm{H}}$
and $L$ as parameters and calculate an emission measure internally. We
then linked the $T_{0}$ and $N_{\mathrm{H}}$ parameters to those of
the \emph{absem} components used in the absorption analysis.

\begin{figure}
  \begin{center}
    \FigureFile(80mm,50mm){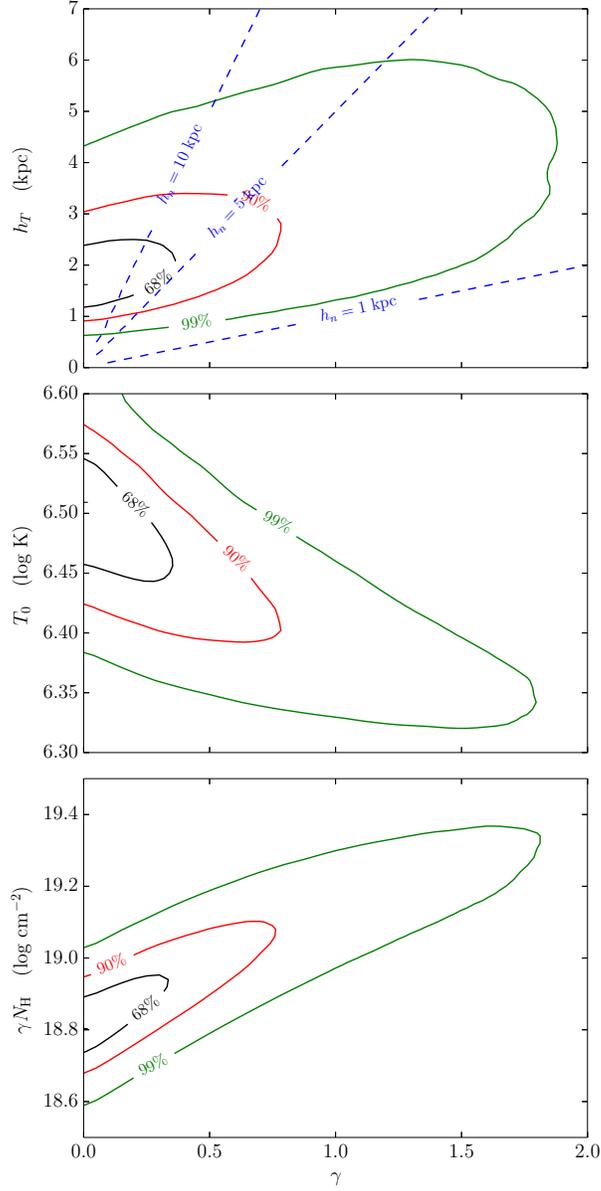}
    \caption{68\%, 90\% and 99\% confidence contours of $h_T$, $T_0$
      and $N_\mathrm{H}$ vs. $\gamma$, obtained from joint fits to the
      X-ray absorption and emission data adopting the exponential
      model. $v_b$ is thawed freely during fits, and the surface
      brightness of LHB+SWCX is fixed to 2 LU. The temperature and
      emission measure of the LTC are fixed to the
      best-fitted values of the emission analysis. In top panel, the
      scale height of the density ($h_n$) is constant along the dashed
      lines.}
    \label{fig:Cont-alpha-hT-T0-NH}
  \end{center}
\end{figure}

\begin{figure}
  \begin{center}
    \FigureFile(80mm,50mm){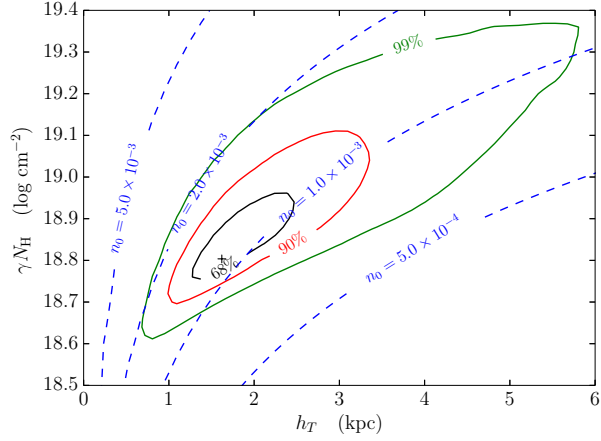}
    \caption{68\%, 90\% and 99\% confidence contours of $h_T$
      vs. $N_\mathrm{H}$, obtained from joint fits to the X-ray
      absorption and emission data adopting the exponential
      model. $v_b$ is thawed freely during fits, and the surface
      brightness of LHB+SWCX is fixed to 2 LU. The temperature and
      emission measure of the LTC are fixed to the
      best-fitted values of the emission analysis. The density at the
      Galactic plane $n_0$ is constant along the dashed lines.}
    \label{fig:Cont-hT-NH}
  \end{center}
\end{figure}

\begin{table*}
  \begin{center}
    \caption{Spectral fitting results in the Exponential model.}
    \label{tab:Fit-Exponential}
    {\small
    \begin{tabular*}{180mm}{@{\extracolsep{\fill}}lccccccccc}
      \hline
      & CXB & \multicolumn{5}{c}{Hot ISM} & \multicolumn{2}{c}{LTC} & $\chi^2$/dof \\ \cline{3-7}
      & Norm\footnotemark[$\dagger$] & $v_b$ & $\log T_0$ & $\log (\gamma N_{\mathrm{H}})$ & $h_T$ & $\gamma$ & $\log T$ & Norm\footnotemark[$\ddagger$] \\
      & & (km s$^{-1}$) & (K) & (cm$^{-2}$) & (kpc) & & (K) \\ \hline
%%% Single PL / Scale = h_T (Xspec 12/fits-pasj8) %%%
      \multicolumn{10}{c}{LHB+SWCX = 2 LU\footnotemark[$\S$]} \\ \hline
      Exp(C) & & & & & & & & & 543.52/518 \\ \cline{1-1}
      \ Abs & $\cdots$ & $59^{+31}_{-12}$ & $6.51^{+0.07}_{-0.12}$ & $18.80^{+0.31}_{-0.14}$ & $1.6^{+1.7}_{-0.7}$ & $0.00(<0.74)$ & $\cdots$ & $\cdots$ \\
      \ Obs1 & $12.1^{+0.5}_{-0.5}$ & $\cdots$ & $\uparrow$ & $\uparrow$ & $\uparrow$ & $\uparrow$ & $\cdots$ & $\cdots$  \\
      \ Obs2(c) & $10.8^{+0.5}_{-0.5}$ & $\cdots$ & $\uparrow$ & $\uparrow$ & $\uparrow$ & $\uparrow$ & 6.59 (fix) & $1.3^{+0.3}_{-0.3}$ \\ \hline
      \multicolumn{4}{@{\extracolsep{\fill}}l}{Exp(C); LTC Lower Limit} & & & & & & 544.53/518 \\ \cline{1-1}
      \ Abs & $\cdots$ & $61^{+30}_{-13}$ & $6.55^{+0.07}_{-0.08}$ & $18.79^{+0.22}_{-0.14}$ & $1.9^{+1.6}_{-0.9}$ & $0.00(<0.46)$ & $\cdots$ & $\cdots$ \\
      \ Obs1 & $12.0^{+0.5}_{-0.5}$ & $\cdots$ & $\uparrow$ & $\uparrow$ & $\uparrow$ & $\uparrow$ & $\cdots$ & $\cdots$  \\
      \ Obs2(c) & $10.8^{+0.5}_{-0.5}$ & $\cdots$ & $\uparrow$ & $\uparrow$ & $\uparrow$ & $\uparrow$ & 6.52 (fix) & $1.5^{+0.4}_{-0.4}$ \\ \hline
      \multicolumn{4}{@{\extracolsep{\fill}}l}{Exp(C); LTC Upper Limit} & & & & & & 547.14/518 \\ \cline{1-1}
      \ Abs & $\cdots$ & $59^{+38}_{-9}$ & $6.50^{+0.07}_{-0.10}$ & $18.80^{+0.18}_{-0.13}$ & $1.4^{+1.6}_{-0.6}$ & $0.00(<0.77)$ & $\cdots$ & $\cdots$ \\
      \ Obs1 & $11.9^{+0.5}_{-0.5}$ & $\cdots$ & $\uparrow$ & $\uparrow$ & $\uparrow$ & $\uparrow$ & $\cdots$ & $\cdots$  \\
      \ Obs2(c) & $10.6^{+0.5}_{-0.5}$ & $\cdots$ & $\uparrow$ & $\uparrow$ & $\uparrow$ & $\uparrow$ & 6.81 (fix) & $0.8^{+0.2}_{-0.2}$ \\ \hline
      \multicolumn{4}{@{\extracolsep{\fill}}l}{Exp(C); Fixed $\gamma$} & & & & & & 547.63/519 \\ \cline{1-1}
      \ Abs & $\cdots$ & $79^{+36}_{-13}$ & $6.39^{+0.07}_{-0.06}$ & $19.08^{+0.04}_{-0.04}$ & $2.2^{+0.8}_{-0.4}$ & 1.0 (fix) & $\cdots$ & $\cdots$ \\
      \ Obs1 & $12.3^{+0.4}_{-0.5}$ & $\cdots$ & $\uparrow$ & $\uparrow$ & $\uparrow$ & $\uparrow$ & $\cdots$ & $\cdots$  \\
      \ Obs2(c) & $10.8^{+0.4}_{-0.5}$ & $\cdots$ & $\uparrow$ & $\uparrow$ & $\uparrow$ & $\uparrow$ & 6.59 (fix) & $1.5^{+0.2}_{-0.3}$ \\ \hline
      \multicolumn{4}{@{\extracolsep{\fill}}l}{Exp(C); Fixed $\gamma$} & & & & & & 548.65/519 \\ \cline{1-1}
      \ Abs & $\cdots$ & $77^{+200}_{-26}$ & $6.36^{+0.05}_{-0.06}$ & $19.25^{+0.17}_{-0.19}$ & $3.5^{+3.5}_{-1.9}$ & 1.5 (fix) & $\cdots$ & $\cdots$ \\
      \ Obs1 & $12.3^{+0.5}_{-0.5}$ & $\cdots$ & $\uparrow$ & $\uparrow$ & $\uparrow$ & $\uparrow$ & $\cdots$ & $\cdots$  \\
      \ Obs2(c) & $10.7^{+0.4}_{-0.5}$ & $\cdots$ & $\uparrow$ & $\uparrow$ & $\uparrow$ & $\uparrow$ & 6.59 (fix) & $1.5^{+0.3}_{-0.2}$ \\ \hline
      \multicolumn{10}{@{}l@{}}{\hbox to 0pt{\parbox{180mm}{\footnotesize
          \par\noindent
          \footnotemark[$\dagger$] In units of $\mathrm{photons\,keV^{-1}\,cm^{-2}\,s^{-1}\,str^{-1}@1keV}$.
          \par\noindent
          \footnotemark[$\ddagger$] In units of $10^{-3}\,\mathrm{pc\,cm^{-6}}$.
          \par\noindent
          \footnotemark[$\S$] Emission measure of \emph{mekal} for LHB+SWCX is set to $0.0066\,\mathrm{pc\,cm^{-6}}$, which corresponds to 2.0 LU O\emissiontype{VII} K$\alpha$ emission.
        }\hss}}
    \end{tabular*}
    }
  \end{center}
\end{table*}

We performed three fits using the best-fitted, lower- and upper-limit
LTC temperatures. In
these fits, $h_n$ tended to be infinity while $\gamma$ tended to be
zero, indicating that the density stays almost uniform while the
temperature drops exponentially away from the plane. For this reason,
we changed the parameters $h_n$ and $N_\mathrm{H}$ of \emph{vabmkl} to
$h_T$ and $\gamma N_\mathrm{H}$ respectively. The fitted results are
shown in Table~\ref{tab:Fit-Exponential}. We obtained the scale
heights for the temperature and density as
$h_{T}=1.6^{+1.7}_{-0.7}\,\mathrm{kpc}$ and $h_{n}>2.8\,\mathrm{kpc}$
and the density at the plane as $n_{0}=(1.2^{+0.5}_{-0.4})\times
10^{-3}\,\mathrm{cm^{-3}}$. Confidence contours of $h_{T}$, $T_{0}$
and $\gamma N_{\mathrm{H}}$ versus $\gamma$, and $\gamma
N_{\mathrm{H}}$ versus $h_{T}$ are shown in
Figure~\ref{fig:Cont-alpha-hT-T0-NH} and Figure~\ref{fig:Cont-hT-NH},
respectively.

Finally, we fixed $\gamma$ to 1 or 1.5 for the scale height ratio
of density and temperature of an ideal gas. The results are also shown
in the same table.

\section{Discussion}

\subsection{Excess in Obs2 spectrum}

\begin{table*}
  \begin{center}
    \caption{Spectral fitting results in the Exponential model fitting the absorption and the emissions with each.}
    \label{tab:Fit-Exponential-Each}
    {\small
    \begin{tabular*}{180mm}{@{\extracolsep{\fill}}lccccccccc}
      \hline
      & CXB & \multicolumn{5}{c}{Hot ISM} & \multicolumn{2}{c}{LTC} & $\chi^2$/dof \\ \cline{3-7}
      & Norm\footnotemark[$\dagger$] & $v_b$ & $\log T_0$ & $\log (\gamma N_{\mathrm{H}})$ & $h_T$ & $\gamma$ & $\log T$ & Norm\footnotemark[$\ddagger$] \\
      & & (km s$^{-1}$) & (K) & (cm$^{-2}$) & (kpc) & & (K) \\ \hline
%%% Single PL (Xspec 12/fits-pasj8) %%%
      \multicolumn{10}{c}{LHB+SWCX = 2 LU\footnotemark[$\S$]} \\ \hline
      \multicolumn{4}{@{\extracolsep{\fill}}l}{With Obs1} & & & & & & 412.65/391 \\ \cline{1-1}
      \ Abs & $\cdots$ & $45^{+14}_{-9}$ & $6.42^{+0.14}_{-0.16}$ & $19.10^{+0.48}_{-0.23}$ & $4.7^{+3.7}_{-2.6}$ & $0.00(<3.25)$ & $\cdots$ & $\cdots$ \\
      \ Obs1 & $12.3^{+0.6}_{-0.5}$ & $\cdots$ & $\uparrow$ & $\uparrow$ & $\uparrow$ & $\uparrow$ & $\cdots$ & $\cdots$ \\ \hline
      \multicolumn{4}{@{\extracolsep{\fill}}l}{With Obs2(a)} & & & & & & 423.9/420 \\ \cline{1-1}
      \ Abs & $\cdots$ & $65^{+31}_{-15}$ & $6.69^{+0.05}_{-0.07}$ & $18.74^{+0.16}_{-0.14}$ & $1.2^{+1.2}_{-0.6}$ & $0.00(<0.14)$ & $\cdots$ & $\cdots$ \\
      \ Obs2(a) & $10.7^{+0.5}_{-0.5}$ & $\cdots$ & $\uparrow$ & $\uparrow$ & $\uparrow$ & $\uparrow$ & $\cdots$ & $\cdots$ \\ \hline
      \multicolumn{4}{@{\extracolsep{\fill}}l}{With Obs2(b)} & & & & & & 421.60/420 \\ \cline{1-1}
      \ Abs & $\cdots$ & $63^{+17}_{-10}$ & $6.62^{+0.06}_{-0.05}$ & $18.76^{+0.13}_{-0.05}$ & $1.1^{+0.5}_{-0.2}$ & $0.00(<0.32)$ & $\cdots$ & $\cdots$ \\
      \ Obs2(b)\footnotemark[$\|$] & $10.7^{+0.5}_{-0.5}$ & $\cdots$ & $\uparrow$ & $\uparrow$ & $\uparrow$ & $\uparrow$ & $\cdots$ & $\cdots$ \\ \hline
      \multicolumn{4}{@{\extracolsep{\fill}}l}{With Obs2(c)} & & & & & & 420.08/419 \\ \cline{1-1}
      \ Abs & $\cdots$ & $61^{+48}_{-9}$ & $6.56^{+0.13}_{-0.13}$ & $18.78^{+0.26}_{-0.13}$ & $1.1^{+1.1}_{-0.5}$ & $0.00(<0.67)$ & $\cdots$ & $\cdots$ \\
      \ Obs2(c) & $10.6^{+0.5}_{-0.5}$ & $\cdots$ & $\uparrow$ & $\uparrow$ & $\uparrow$ & $\uparrow$ & 6.83 (fix) & $0.5^{+0.4}_{-0.5}$ \\ \hline
      \multicolumn{10}{@{}l@{}}{\hbox to 0pt{\parbox{180mm}{\footnotesize
          \par\noindent
          \footnotemark[$\dagger$] In units of $\mathrm{photons\,keV^{-1}\,cm^{-2}\,s^{-1}\,str^{-1}@1keV}$.
          \par\noindent
          \footnotemark[$\ddagger$] In units of $10^{-3}\,\mathrm{pc\,cm^{-6}}$.
          \par\noindent
          \footnotemark[$\S$] Emission measure of \emph{mekal} for LHB+SWCX is set to $0.0033\,\mathrm{pc\,cm^{-6}}$, which corresponds to 1.0 LU O\emissiontype{VII} K$\alpha$ emission as the estimated lower limit.
          \par\noindent
          \footnotemark[$\|$] Abundances of Ne and Fe are set to $2.1$ and $1.2$ respectively, according to the Obs2(c) fit in Table~\ref{tab:Emi-Fit-Results}.
        }\hss}}
    \end{tabular*}
    }
  \end{center}
\end{table*}

There was a large discrepancy in the spectra extracted from Obs1 and
Obs2. The spectrum of Obs2 shows the excess in the energy range from
0.6 to 1.0 keV.  \citet{yoshino09} observed twelve blank fields with
\emph{Suzaku} and found similar excesses in four of them. As the
excesses, seen in one third of the observations, are not
extraordinary. The excess seen Obs2 may be rare because it is just a
degree away from Obs1. \citet{yao09}, \citet{hagihara10} and this
paper assumed that emissions from two \emph{Suzaku} observations
1$^\circ$ apart on the opposite sides of the respective X-ray sources
are the same. But this may be too simplistic. To estimate the effect
of the fluctuation, we jointly fitted the absorption spectrum with
each emission spectra. The results in
Table~\ref{tab:Fit-Exponential-Each} indicate that this degree of
fluctuation does not affect the characteristic of the hot ISM, as
inferred from our analysis.

What could be the cause of such fluctuation? The temperature of the
excess seems too high to be a fluctuation in the LHB.
It may be due to the cosmic variance (extragalactic background
fluctuation due to the large-scale structure): e.g., due to the
presence of a distant group/cluster of galaxies or two, which are
too faint to be detected individually in the observation.
The high-velocity clouds (HVCs) could be another candidate. Mkn 421 is
projected inside the HVC Complex M. It is considered to be near
the Galactic disk \citep{blitz99}, and may thus interact strongly with
the hot ISM. The origin is still unknown because neither the distance
nor the apparent size are known.

\subsection{Uncertainty due to the SWCX and LHB}\label{sec:uncertainty-due-swcx}

\begin{table*}
  \begin{center}
    \caption{Spectral fitting results of emission data using the lower and upper limits of SWCX and LHB.}
    \label{tab:Emi-Fit-Results-SWCX}
    {\small
    \begin{tabular*}{180mm}{@{\extracolsep{\fill}}lcccccccccc}
      \hline
      & CXB & \multicolumn{2}{c}{LHB+SWCX} & \multicolumn{4}{c}{Hot ISM} & \multicolumn{2}{c}{LTC} & $\chi^2$/dof \\ \cline{5-8}
      & Norm\footnotemark[$\dagger$] & $\log T$ (K) & Norm\footnotemark[$\ddagger$] & $\log T$ (K) & Norm\footnotemark[$\ddagger$] & Ne/O & Fe/O & $\log T$ (K) & Norm\footnotemark[$\ddagger$] \\ \hline
%%% Single PL (Xspec 12/fits-pasj8) %%%
      \multicolumn{11}{c}{LHB+SWCX = 1 LU\footnotemark[$\S$] (Lower Limit)} \\ \hline
      \multicolumn{2}{@{\extracolsep{\fill}}l}{Emission(C)} & & & & & & & & & 262.89/223 \\ \cline{1-1}
      \ Obs1 & $12.2^{+0.5}_{-0.5}$ & 6.06 (fix) & 3.3 (fix) & $6.31^{+0.02}_{-0.06}$ & $3.2^{+0.7}_{-0.4}$ & 1 (fix) & 1 (fix) & $\cdots$ & $\cdots$ \\
      \ Obs2(c) & $10.7^{+0.5}_{-0.5}$ & $\uparrow$ & $\uparrow$ & $\uparrow$ & $\uparrow$ & $\uparrow$ & $\uparrow$ & $6.63^{+0.18}_{-0.10}$ & $1.3^{+0.2}_{-0.4}$ \\ \hline
      \multicolumn{11}{c}{LHB+SWCX = 3 LU\footnotemark[$\|$] (Upper Limit)} \\ \hline
      \multicolumn{2}{@{\extracolsep{\fill}}l}{Emission(C)} & & & & & & & & & 259.21/223 \\ \cline{1-1}
      \ Obs1 & $12.1^{+0.6}_{-0.5}$ & 6.06 (fix) & 9.8 (fix) & $6.43^{+0.10}_{-0.10}$ & $1.25^{+0.75}_{-0.42}$ & 1 (fix) & 1 (fix) & $\cdots$ & $\cdots$ \\
      \ Obs2(c) & $10.8^{+0.5}_{-0.5}$ & $\uparrow$ & $\uparrow$ & $\uparrow$ & $\uparrow$ & $\uparrow$ & $\uparrow$ & $6.55^{+0.23}_{-0.09}$ & $1.5^{+0.6}_{-0.5}$ \\ \hline
      \multicolumn{11}{@{}l@{}}{\hbox to 0pt{\parbox{180mm}{\footnotesize
          \par\noindent
          \footnotemark[$\dagger$] In units of $\mathrm{photons\,keV^{-1}\,cm^{-2}\,s^{-1}\,str^{-1}@1keV}$.
          \par\noindent
          \footnotemark[$\ddagger$] In units of $10^{-3}\,\mathrm{pc\,cm^{-6}}$.
          \par\noindent
          \footnotemark[$\S$] The emission measure of LHB+SWCX is fixed to $0.0033\,\mathrm{pc\,cm^{-6}}$ which corresponds to 1.0 LU of O\emissiontype{VII} K$\alpha$ emission as the estimated lower limit.
          \par\noindent
          \footnotemark[$\|$] The emission measure of LHB+SWCX is fixed to $0.0098\,\mathrm{pc\,cm^{-6}}$ which corresponds to 3.0 LU of O\emissiontype{VII} K$\alpha$ emission as the estimated lower limit.
      }\hss}}
    \end{tabular*}
    }
  \end{center}
\end{table*}

\begin{table*}
  \begin{center}
    \caption{Results from the spectral fits with the exponential
      model, together with the lower and upper limits to the SWCX and
      LHB contributions.}
    \label{tab:Fit-Exponential-SWCX}
    {\small
    \begin{tabular*}{180mm}{@{\extracolsep{\fill}}lccccccccc}
      \hline
      & CXB & \multicolumn{5}{c}{Hot ISM} & \multicolumn{2}{c}{LTC} & $\chi^2$/dof \\ \cline{3-7}
      & Norm\footnotemark[$\dagger$] & $v_b$ & $\log T_0$ & $\log (\gamma N_{\mathrm{H}})$ & $h_T$ & $\gamma$ & $\log T$ & Norm\footnotemark[$\ddagger$] \\
      & & (km s$^{-1}$) & (K) & (cm$^{-2}$) & (kpc) & & (K) \\ \hline
%%% Single PL (Xspec 12/fits-pasj8) %%%
%% 1LU
      \multicolumn{10}{c}{LHB+SWCX = 1 LU\footnotemark[$\S$] (Lower Limit)} \\ \hline
      Exp(C) & & & & & & & & & 545.77/518 \\ \cline{1-1}
      \ Abs & $\cdots$ & $59^{+35}_{-13}$ & $6.41^{+0.09}_{-0.08}$ & $19.00^{+0.34}_{-0.24}$ & $1.6^{+2.2}_{-0.9}$ & $0.39(<1.60)$ & $\cdots$ & $\cdots$ \\
      \ Obs1 & $12.2^{+0.5}_{-0.6}$ & $\cdots$ & $\uparrow$ & $\uparrow$ & $\uparrow$ & $\uparrow$ & $\cdots$ & $\cdots$  \\
      \ Obs2(c) & $10.7^{+0.5}_{-0.5}$ & $\cdots$ & $\uparrow$ & $\uparrow$ & $\uparrow$ & $\uparrow$ & 6.63 (fix) & $1.3^{+0.3}_{-0.3}$ \\ \hline
      \multicolumn{4}{@{\extracolsep{\fill}}l}{Exp(C); LTC Lower Limit} & & & & & & 547.42/518 \\ \cline{1-1}
      \ Abs & $\cdots$ & $57^{+28}_{-8}$ & $6.47^{+0.06}_{-0.14}$ & $18.83^{+0.47}_{-0.14}$ & $1.3^{+2.3}_{-0.6}$ & $0.00(<1.42)$ & $\cdots$ & $\cdots$ \\
      \ Obs1 & $12.1^{+0.5}_{-0.5}$ & $\cdots$ & $\uparrow$ & $\uparrow$ & $\uparrow$ & $\uparrow$ & $\cdots$ & $\cdots$  \\
      \ Obs2(c) & $10.8^{+0.4}_{-0.4}$ & $\cdots$ & $\uparrow$ & $\uparrow$ & $\uparrow$ & $\uparrow$ & 6.53 (fix) & $1.6^{+0.4}_{-0.4}$ \\ \hline
      \multicolumn{4}{@{\extracolsep{\fill}}l}{Exp(C); LTC Upper Limit} & & & & & & 548.75/518 \\ \cline{1-1}
      \ Abs & $\cdots$ & $58^{+20}_{-11}$ & $6.45^{+0.05}_{-0.06}$ & $18.88^{+0.34}_{-0.16}$ & $1.2^{+1.0}_{-0.6}$ & $0.11(<0.77)$ & $\cdots$ & $\cdots$ \\
      \ Obs1 & $12.0^{+0.5}_{-0.5}$ & $\cdots$ & $\uparrow$ & $\uparrow$ & $\uparrow$ & $\uparrow$ & $\cdots$ & $\cdots$  \\
      \ Obs2(c) & $10.6^{+0.5}_{-0.5}$ & $\cdots$ & $\uparrow$ & $\uparrow$ & $\uparrow$ & $\uparrow$ & 6.81 (fix) & $0.9^{+0.2}_{-0.2}$ \\ \hline
%% 3LU
      \multicolumn{10}{c}{LHB+SWCX = 3 LU\footnotemark[$\|$] (Upper Limit)} \\ \hline
      Exp(C) & & & & & & & & & 546.51/518 \\ \cline{1-1}
      \ Abs & $\cdots$ & $57^{+29}_{-9}$ & $6.46^{+0.03}_{-0.11}$ & $18.85^{+0.48}_{-0.14}$ & $1.4^{+2.6}_{-0.6}$ & $0.02 (<1.60)$ & $\cdots$ & $\cdots$ \\
      \ Obs1 & $12.1^{+0.5}_{-0.5}$ & $\cdots$ & $\uparrow$ & $\uparrow$ & $\uparrow$ & $\uparrow$ & $\cdots$ & $\cdots$  \\
      \ Obs2(c) & $10.8^{+0.5}_{-0.5}$ & $\cdots$ & $\uparrow$ & $\uparrow$ & $\uparrow$ & $\uparrow$ & 6.55 (fix) & $1.5^{+0.4}_{-0.3}$ \\ \hline
      \multicolumn{4}{@{\extracolsep{\fill}}l}{Exp(C); LTC Lower Limit} & & & & & & 553.45/518 \\ \cline{1-1}
      \ Abs & $\cdots$ & $62^{+24}_{-15}$ & $6.53^{+0.07}_{-0.09}$ & $18.77^{+0.23}_{-0.12}$ & $1.4^{+1.5}_{-0.8}$ & $0.00(<0.46)$ & $\cdots$ & $\cdots$ \\
      \ Obs1 & $12.0^{+0.6}_{-0.5}$ & $\cdots$ & $\uparrow$ & $\uparrow$ & $\uparrow$ & $\uparrow$ & $\cdots$ & $\cdots$  \\
      \ Obs2(c) & $10.9^{+0.5}_{-0.5}$ & $\cdots$ & $\uparrow$ & $\uparrow$ & $\uparrow$ & $\uparrow$ & 6.46 (fix) & $1.7^{+0.5}_{-0.4}$ \\ \hline
      \multicolumn{4}{@{\extracolsep{\fill}}l}{Exp(C); LTC Upper Limit} & & & & & & 547.36/518 \\ \cline{1-1}
      \ Abs & $\cdots$ & $59^{+36}_{-13}$ & $6.44^{+0.06}_{-0.12}$ & $18.90^{+0.34}_{-0.18}$ & $1.2^{+1.7}_{-0.6}$ & $0.15(<1.23)$ & $\cdots$ & $\cdots$ \\
      \ Obs1 & $12.0^{+0.7}_{-0.5}$ & $\cdots$ & $\uparrow$ & $\uparrow$ & $\uparrow$ & $\uparrow$ & $\cdots$ & $\cdots$  \\
      \ Obs2(c) & $10.6^{+0.2}_{-0.5}$ & $\cdots$ & $\uparrow$ & $\uparrow$ & $\uparrow$ & $\uparrow$ & 6.77 (fix) & $0.9^{+0.2}_{-0.2}$ \\ \hline
      \multicolumn{10}{@{}l@{}}{\hbox to 0pt{\parbox{180mm}{\footnotesize
          \par\noindent
          \footnotemark[$\dagger$] In units of $\mathrm{photons\,keV^{-1}\,cm^{-2}\,s^{-1}\,str^{-1}@1keV}$.
          \par\noindent
          \footnotemark[$\ddagger$] In units of $10^{-3}\,\mathrm{pc\,cm^{-6}}$.
          \par\noindent
          \footnotemark[$\S$] Emission measure of \emph{mekal} for LHB+SWCX is set to $0.0033\,\mathrm{pc\,cm^{-6}}$, which corresponds to 1.0 LU O\emissiontype{VII} K$\alpha$ emission as the estimated lower limit.
          \par\noindent
          \footnotemark[$\|$] Emission measure of \emph{mekal} for LHB+SWCX is set to $0.0098\,\mathrm{pc\,cm^{-6}}$, which corresponds to 3.0 LU O\emissiontype{VII} K$\alpha$ emission as the estimated upper limit.
        }\hss}}
    \end{tabular*}
    }
  \end{center}
\end{table*}

The contamination from the geocoronal SWCX can be eliminated by
screening the event data properly, but the heliospheric SWCX and LHB
can not. Shadowing measurements show that the heliospheric SWCX and
LHB contribute almost negligibly to the total above 500~eV, and their
only effect on our analysis is their somewhat uncertain contribution
of $2\pm 1$~LU to the measured O\emissiontype{VII} emission
(\cite{yoshino09}). We also performed a model simulation, which we
developed for \citet{yoshitake13}, accroding to \citet{kou06} and
obtained the line intensities contributed from the heliospheric SWCX
as 1.9~LU and 2.2~LU for solar minimum and solar maximum respectively,
which are consistent with our estimation.  We thus modeled the
heliospheric SWCX and LHB using the parameter-fixed \emph{mekal} to
represent their O\emissiontype{VII} emission line.

We first obtained the best-fit and the lower and upper limits of LTC
from the emission analyses using the lower and upper limits of the
heliospheric SWCX and LHB O\emissiontype{VII} emission
(Table~\ref{tab:Emi-Fit-Results-SWCX}). We then applied the obtained
LTC temperatures to the joint fits (Table~\ref{tab:Fit-Exponential-SWCX}).

The results from the fits using the lower and upper limits are
consistent with those from the fits using the typical line intensity
for the heliospheric SWCX and LHB contributions.

\subsection{Distribution of the O\emissiontype{VII} and O\emissiontype{VIII} Emitting and Absorbing Gas}

\begin{figure}
  \begin{center}
    \FigureFile(80mm,50mm){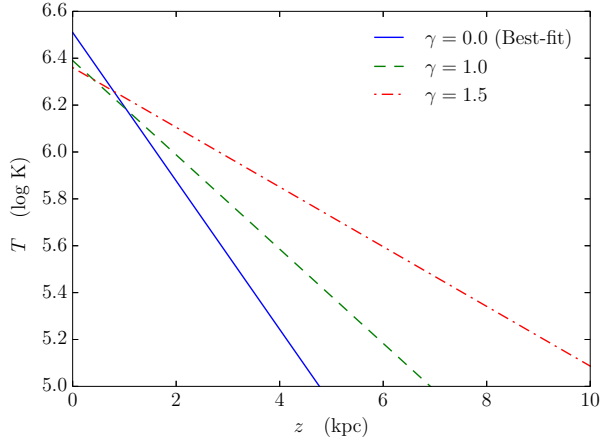}
    \caption{Temperature drop along with the distance from the
      Galactic plane, predicted with the exponential model with the
      fixed LTC temperature and the thawed $v_b$, for the
      best-fitted $\gamma$, as well as for the unity scale height of
      density and temperature ($\gamma = 1.0$) and the ideal gas
      ($\gamma = 1.5$).}
  \label{fig:Temperature-Drop}
  \end{center}
\end{figure}

\begin{figure}
  \begin{center}
    \FigureFile(80mm,50mm){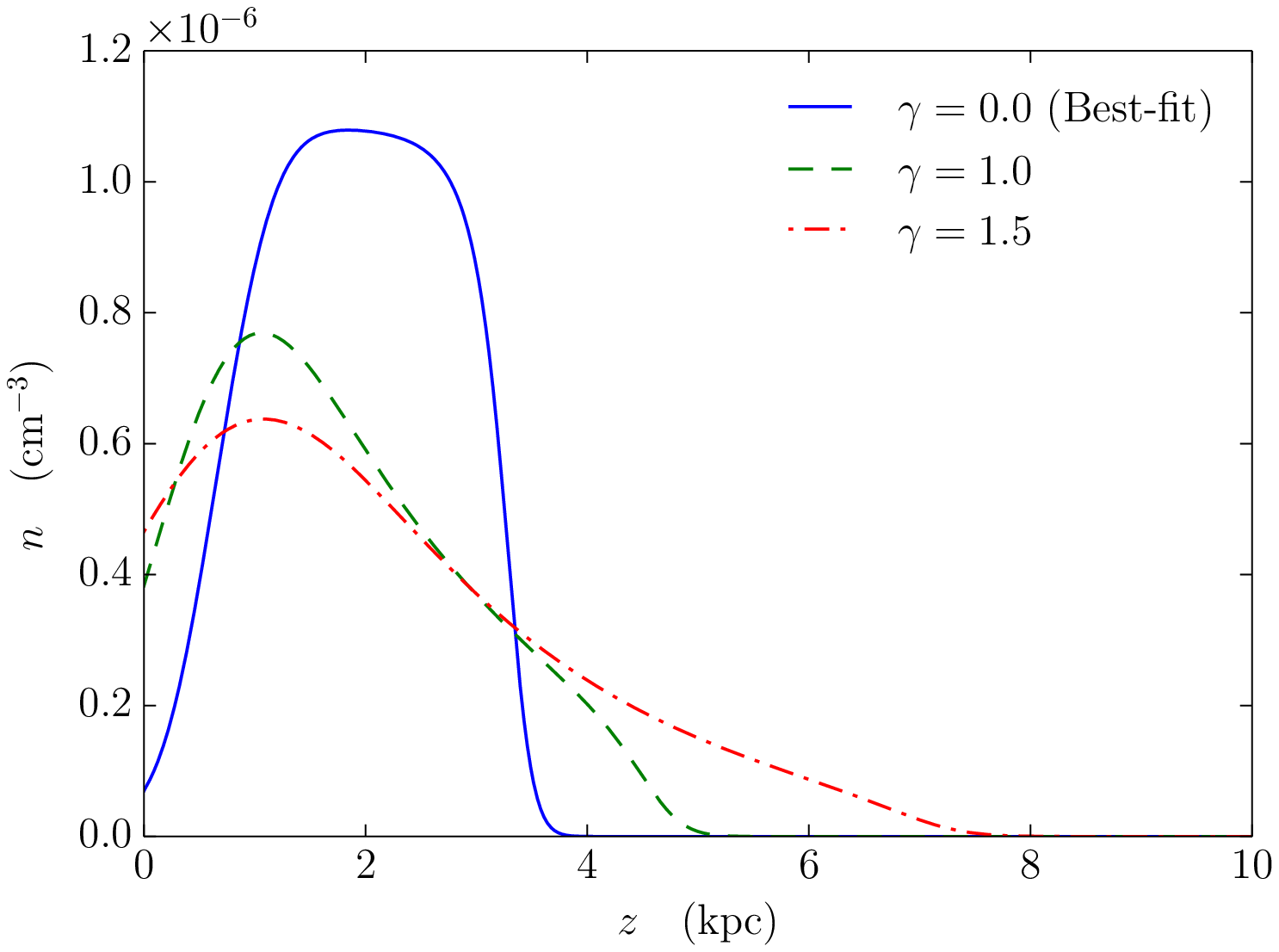} \\
    \FigureFile(80mm,50mm){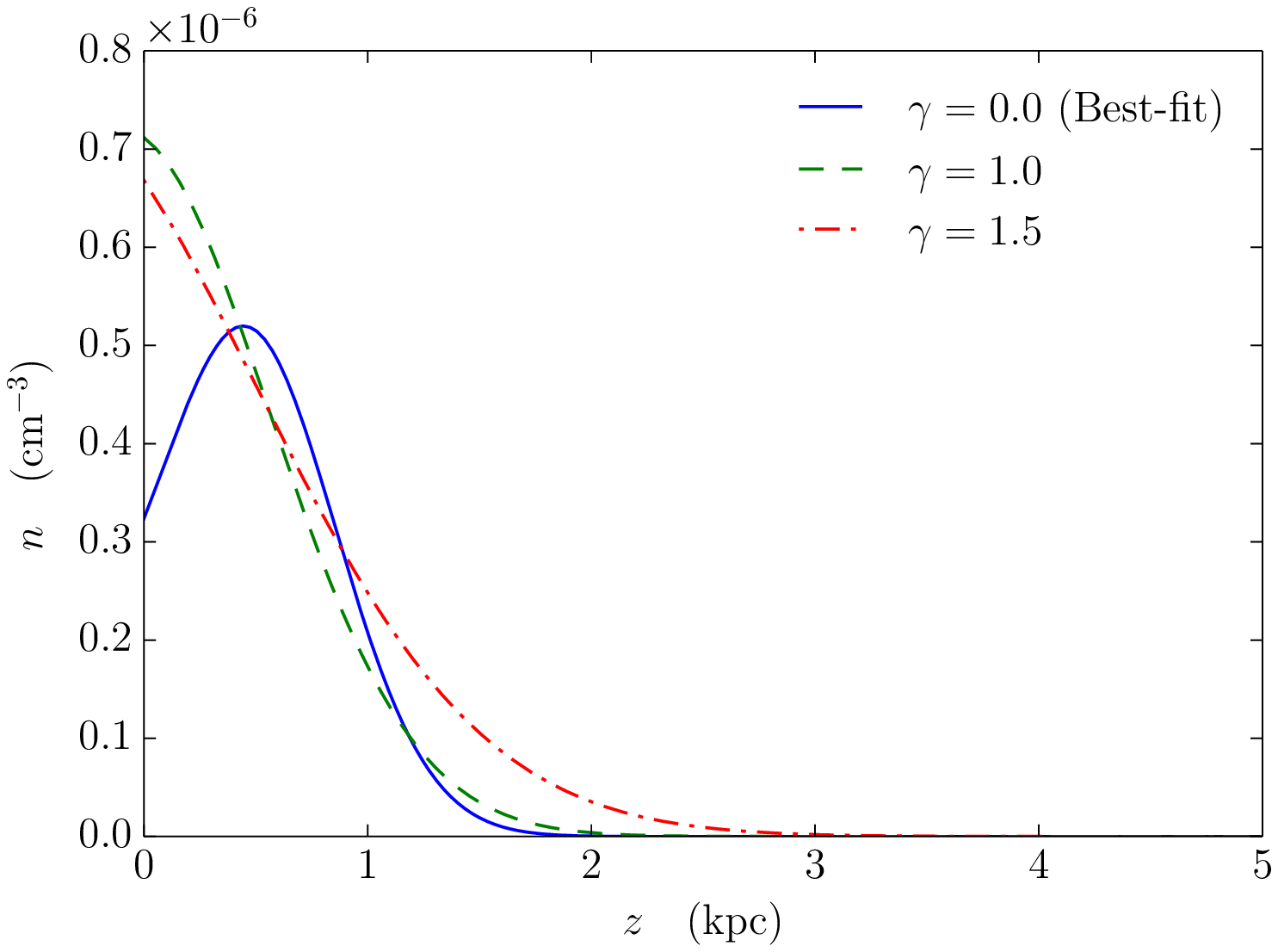}
    \caption{Spatial distribution of the O\emissiontype{VII} (top) and
      O\emissiontype{VIII} (bottom) density as the functions of the
      distance from the Galactic plane, predicted with the exponential
      model, for the best-fitted and the fixed $\gamma$-s with the
      fixed LTC temperature and the thawed $v_b$, for the best-fitted
      $\gamma$, as well as for the unity scale height of density and
      temperature ($\gamma = 1.0$) and the ideal gas ($\gamma =
      1.5$).}
  \label{fig:Oxygen-Column-Density}
  \end{center}
\end{figure}

\begin{figure}
  \begin{center}
    \FigureFile(80mm,50mm){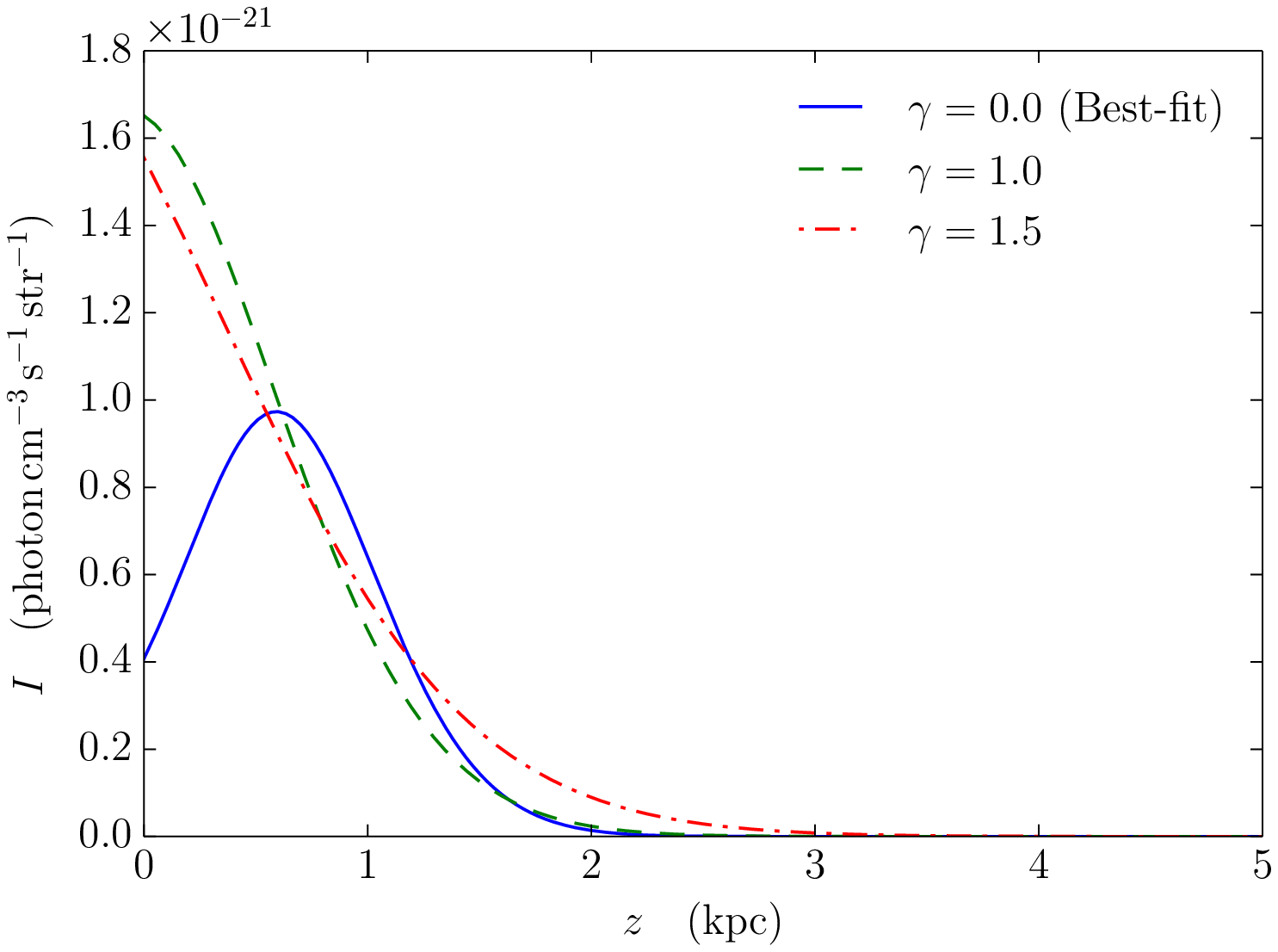} \\
    \FigureFile(80mm,50mm){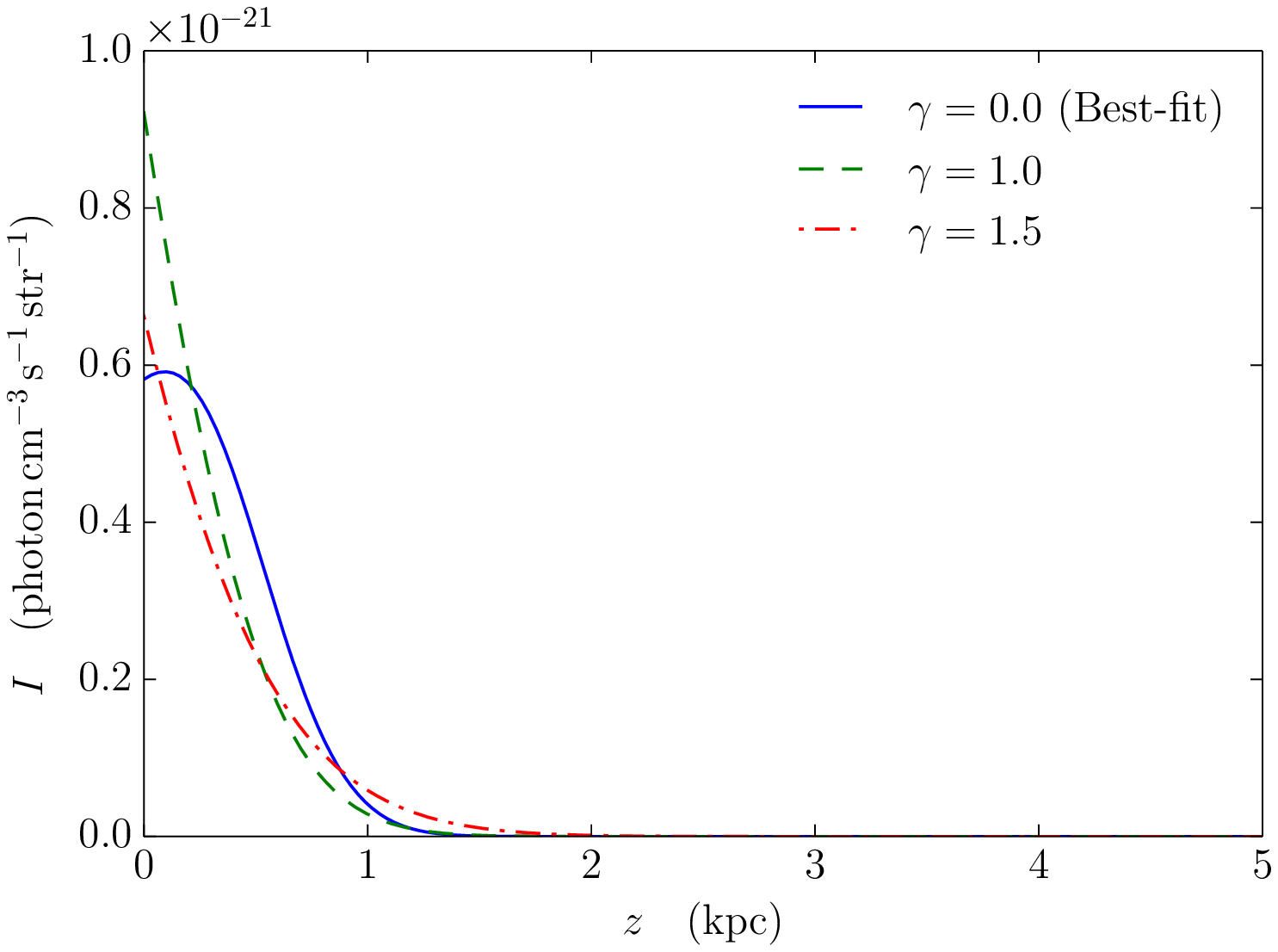}
    \caption{Spatial distribution of the O\emissiontype{VII} (top) and
      O\emissiontype{VIII} (bottom) surface brightness as the
      functions of the distance from the Galactic plane, predicted
      with the exponential model with the fixed LTC temperature and
      the thawed $v_b$, for the best-fitted $\gamma$, as well as for
      the unity scale height of density and temperature ($\gamma =
      1.0$) and the ideal gas ($\gamma = 1.5$).}
  \label{fig:Oxygen-Surface-Brightness}
  \end{center}
\end{figure}

Assuming the exponential model, we obtained the temperature and
hydrogen number density at the Galactic plane, as well as the scale
heights. Figure~\ref{fig:Temperature-Drop} shows the temperature as a
function of the distance from the Galactic plane under the
best-fitted and the fixed $\gamma$-s (1 or 1.5 for the scale height
ratio of density and temperature of an ideal gas).
In any $\gamma$-s, the fitted results are consistent and the temperature
scale heights take finite values less than 10~kpc.
The velocity dispersions are also consistent with the thermal velocities
($\sim 60\,\mathrm{km/s}$).

Using the fitted parameters, we calculated the oxygen number density
(Figure~\ref{fig:Oxygen-Column-Density}) and the surface brightness
(Figure~\ref{fig:Oxygen-Surface-Brightness}) for O\emissiontype{VII}
and O\emissiontype{VIII} emission lines as functions of the distance
from the Galactic plane. We further computed the vertical heights at
which the accumulated O\emissiontype{VII} and O\emissiontype{VIII}
column density and surface brightness reach $90\%$ of their totals
(Table~\ref{tab:90-CD-and-SB}). In all cases, the obtained heights
are less than 5 kpc.

\begin{table}
  \begin{center}
    \caption{90\% heights calculated from the density and the emission
      measure of the hot gas.}
    \label{tab:90-CD-and-SB}
    \begin{tabular}{lcccc}
      \hline
      & \multicolumn{2}{c}{Absorption} & \multicolumn{2}{c}{Emission} \\
      & O\emissiontype{VII} & O\emissiontype{VIII} & O\emissiontype{VII} & O\emissiontype{VIII} \\
      & \multicolumn{2}{c}{(kpc)} & \multicolumn{2}{c}{(kpc)} \\ \hline
      $\gamma=0.0$ (Best-fit) & 2.98 & 1.00 & 1.23 & 0.70 \\
      $\gamma=1.5$ (Fixed) & 4.70 & 1.39 & 1.43 & 0.84 \\
      $\gamma=1.0$ (Fixed) & 3.50 & 1.01 & 1.11 & 0.63 \\ \hline
    \end{tabular}
  \end{center}
\end{table}

\subsection{Comparison with Other Works}

\citet{yao09} and \citet{hagihara10} performed joint analyses of the
absorption and emission spectra in the LMC X--3 and PKS 2155-304
directions and constrained the hot gas temperatures and densities at
the plane together with the scale heights.

Here we compared our results with those directions
(Table~\ref{tab:Results-Comparison}). The scale height $h$ in the
table represents the scale height for the temperature $h_T$ for LMC
X--3 and PKS 2155--304, and the scale height for the density $h_n$ for
Mkn 421, as we changed the parameter $h_n$ of \emph{vabmkl} to
$h_T$. These values indicate a thick hot gas disk surrounding our
Galaxy, which can be characterized by the exponential model of the
column density, scale height and temperature: $\sim 2\times
10^{19}~\mathrm{cm^{-2}}$, a few kpc and $\sim 2\times
10^6~\mathrm{K}$, respectively.

\citet{Gupta12} combined an averaged column density from the
\emph{Chandra} absorption spectra with an averaged emission measure
from literature and suggested that the spatial extent of the hot gas
is more than 100~kpc.  Moreover, \citet{Gupta13} combined the
\emph{Chandra} absorption spectra toward Mkn 421 with the same
\emph{Suzaku} emission spectra as we used in this paper, and claimed
that the spatial extent and the density of the hot gas toward Mkn 421
are $334^{+685}_{-274}$ kpc and $(1.6^{+2.6}_{-0.8})\times
10^{-4}\,\mathrm{cm^{-3}}$, which differ by more than an order of
magnitude from our results. These differences mainly come from the gas
distribution model and the oxygen abundance.  They assumed a uniformly
distributed isothermal gas with 0.2~$Z_\odot$, while we assumed the
exponentially distributed gas with 1~$Z_\odot$.  Since $EM\propto Z
n^2 L$ and $CD\propto Z n L$, $L\propto CD^2/(Z\,EM)$, the difference
in abundance therefore makes their spatial extent five time larger
than our result.  The difference in the distribution model also makes
the extent differ.  Since the gas temperatures obtained from the
absorption line analysis and the emission line analysis are
inconsistent, as with previous studies in two other directions, a
model with temperature variation along the vertical distance from the
Galactic plane is more appropriate than an isothermal model.  In this
paper, we adopted the exponential model that the temperature and
density decrease exponentially with the vertical distance from the
Galactic plane.  As for the hot gas observable by highly ionized
oxygen, our results are consistent with observations of highly ionized
absorption lines in spectra of Galactic sources (e.g., \cite{fut04})
indicating that most of hot gas contributing to the absorption exists
within distance scales of $\sim 10$~kpc.

\begin{table*}
  \begin{center}
    \caption{Obtained exponential model parameters with LMC X--3 and PKS 2155--304 sight lines.}    
    \label{tab:Results-Comparison}
    \begin{tabular}{lccccc}
      \hline
      Direction & $(\ell, b)$ & $\log T_0$ & $\log N_{\mathrm{H_{Hot}}}$ & $h$\footnotemark[$\dagger$] & $\gamma$ \\
      & ($^\circ$) & (K) & (cm$^{-2}$) & (kpc) \\ \hline
      Mkn 421 & $(179.8, 65.0)$ & $6.51^{+0.07}_{-0.12}$ & $18.80^{+0.31}_{-0.14}$ & $1.6^{+1.7}_{-0.7}$ & $0.00(<0.74)$ \\
      LMC X--3 & $(273.6, -32.1)$ & $6.56^{+0.11}_{-0.10}$ & $19.36^{+0.22}_{-0.21}$ & $2.8^{+3.6}_{-1.8}$ & $0.5^{+1.2}_{-0.4}$ \\
      PKS 2155--304 & $(17.7, -52.2)$ & $6.40^{+0.09}_{-0.05}$ & $19.10^{+0.08}_{-0.07}$ & $2.3^{+0.9}_{-0.8} $ & $2.44^{+1.11}_{-1.41}$ \\ \hline
      \multicolumn{6}{@{}l@{}}{\hbox to 0pt{\parbox{180mm}{\footnotesize
          \par\noindent
          \footnotemark[$\dagger$] $h_n$ for LMC X--3 and PKS 2155--304, and $h_T$ for Mkn 421.
        }\hss}}
      
    \end{tabular}
  \end{center}
\end{table*}

\section{Summary}

We have jointly analyzed X-ray absorption and emission spectral
observed by \emph{Chandra} and \emph{Suzaku} to study the structure of
the hot ISM along the sight line toward Mkn 421. Our main results and
conclusions are summarized as follows:

\begin{enumerate}
\item We have detected O\emissiontype{VII} and O\emissiontype{VIII}
  lines in the emission spectra from two \emph{Suzaku} observations
  taken in the fields adjacent to the Mkn 421 sight line. The two
  spectra show significant difference, indicating a patchy
  thermal emission component to Obs2 spectrum. By modeling these
  emission spectra (excluding local contributions), we obtained a
  characteristic temperature of the emitting gas as
  $(2.0^{+0.2}_{-0.3})\times 10^{6}\,\mathrm{K}$, assumed to be
  isothermal and in the CIE. This temperature is about 30\% higher
  than that of the absorption gas under the same assumption.
\item We have jointly analyzed the \emph{Chandra} absorption and
  \emph{Suzaku} emission spectra adopting the exponential thick
  Galactic hot gas disk model, and found that obtained the gas
  temperature and the density to be $(3.2^{+0.6}_{-0.7})\times 10^{6}\,\mathrm{K}$ and
  $(1.2^{+0.5}_{-0.4})\times 10^{-3}\,\mathrm{cm^{-3}}$ at the Galactic plane
  and have the scale heights of $1.6^{+1.7}_{-0.7}\,\mathrm{kpc}$ and
  $>2.8\,\mathrm{kpc}$, respectively.
\item The results we have obtained from these joint analysis are
  consistent with those for the LMC X--3 and PKS 2155--304 directions,
  consistent with the presence of the thick hot gas disk
  with a vertical scale height of a few kpc.
\end{enumerate}

\bigskip

YY acknowledges financial support by NASA through ADP grant NNH12CG14C
to Eureka Scientific, while QDW is supported by NASA via grant
NNX10AE85G.

\end{document}